\documentclass[preprint2]{aastex}

\usepackage{amsmath}
\usepackage{amssymb}
\usepackage{fancyhdr}
\usepackage{graphicx}
\usepackage{subfigure}
\usepackage{afterpage}
\usepackage{pifont}
\newcommand{\tick}{\ding{52}}

\newcommand{\moo}{\rm $\mu$m}

\shorttitle{Spitzer spectroscopy of the 2Jy and 3CRR samples}
\shortauthors{Dicken et al.}

\begin{document}

\title{Spitzer mid-IR spectroscopy of powerful 2Jy and 3CRR radio galaxies. I. Evidence against a strong starburst-AGN connection in radio-loud AGN}

\author{D. Dicken\altaffilmark{1},
 C. Tadhunter\altaffilmark{2},
 D. Axon\altaffilmark{1,3},
 R. Morganti\altaffilmark{4,5}, 
 A. Robinson\altaffilmark{1}, 
 M.B.N. Kouwenhoven\altaffilmark{6,7},
 H. Spoon\altaffilmark{8},
 P. Kharb\altaffilmark{1},
 K. J. Inskip\altaffilmark{9},
 J. Holt\altaffilmark{10},
 C. Ramos Almeida\altaffilmark{2},
 N. P. H. Nesvadba\altaffilmark{11}
}

    \altaffiltext{1}{Rochester
    Institute of Technology, 84 Lomb Memorial Drive, Rochester NY
    14623, USA ; daniel.dicken@ias.u-psud.fr}
    \altaffiltext{2}{University of
    Sheffield, Hounsfield Road, Sheffield, S3 7RH, UK;} 
    \altaffiltext{3}{University of
    Sussex, Pevensey 2, University of Sussex, Falmer, Brighton, BN1 9QH, UK} 
    \altaffiltext{4}{ASTRON, P.O. Box 2,
    7990 AA Dwingeloo The Netherlands}
    \altaffiltext{5}{Kapetyn Astronmical Institute, University of Groningen, Postbuss 800, 9700 AV 
Groningen, The netherlands}
   \altaffiltext{6}{Kavli Institute for Astronomy and Astrophysics, Peking University, Yi He Yuan Lu 5, Haidian Qu, Beijing 100871, China} 
   \altaffiltext{7}{IAU Peter and Patricia Gruber Foundation Fellow} 
    \altaffiltext{8}{224 Space Sciences Building, Cornell University, Ithaca, NY 14853} 
    \altaffiltext{9}{Max Planck Institute for Astronomy, K\"onigstuhl 17, 69117 Heidelberg, Germany}
\altaffiltext{10}{Leiden Observatory, Leiden University, PO Box 9513, 2300 RA Leiden, the Netherlands}
\altaffiltext{11}{Institut dÕAstrophysique Spatiale, CNRS, Universit\'e Paris Sud, 91405 Orsay, France}

\begin{abstract}
We present deep Spitzer/IRS spectra for complete samples of 46  2Jy radio galaxies (0.05$<$z$<$0.7) and 19 3CRR FRII radio galaxies (z$<$0.1), and use the detection of polycyclic aromatic hydrocarbon (PAH) features to examine the incidence of contemporaneous star formation and radio-loud AGN activity. Our analysis reveals PAH features in only a minority (30\%) of the objects with good IRS spectra. Using the wealth of complementary data available for the 2Jy and 3CRR samples we  make  detailed comparisons between a range of star formation diagnostics: optical continuum spectroscopy, mid- to far-IR (MFIR) color, far-IR excess and PAH detection. There is good agreement between the various diagnostic techniques: most candidates identified to have star formation activity on the basis of PAH detection are also identified using at least two of the other techniques. We find that only 35\% of the combined 2Jy and 3CRR sample show evidence for recent star formation activity (RSFA) at optical and/or MFIR wavelengths. This result argues strongly against the idea of a close link between starburst and powerful radio-loud AGN activity, reinforcing the view that, although a large fraction of powerful radio galaxies may be triggered in galaxy interactions, only a minority are triggered at the peaks of star formation activity in major, gas-rich mergers. However, we find that compact radio sources ($D < 15$~kpc) show a significantly higher incidence of RSFA ($>$75\%) than their more extended counterparts ($\approx$15 -- 25\%). We discuss this result in the context of a possible bias towards the selection of compact radio sources triggered in gas-rich environments.
\end{abstract}


\section{Introduction}
\label{sec:intro}

Extragalactic radio sources form an important subset of the Active Galactic Nuclei (AGN) population because their relativistic jets heat the Inter-Stellar/Galactic Medium (ISM/IGM) and drive outflows on scales ranging from the kpc-scale narrow line regions \citep{holt08}, to the 100~kpc -- 1~Mpc hot X-ray haloes of the host galaxies and galaxy clusters \citep{mcnamara07}. Such feedback has been used to explain the cooling problem for the hot X-ray gas \citep{best05}, as well as the high end shape of the galaxy luminosity function \citep{benson03}, and the correlations between black hole mass and host galaxy properties (\citealp{bower06,croton06}). Therefore, given that a large percentage of giant elliptical galaxies may go through a radio-loud phase \citep{best06}, the triggering of radio galaxies is a key issue for the understanding of the evolution of massive galaxies in general. However, there remain considerable uncertainties about how and when radio-loud AGN activity is triggered.

Morphological and spectroscopic studies provide strong evidence that powerful radio galaxies are triggered in galaxy mergers, although not necessarily at a single phase of a particular type of merger \citep{heckman86,ramos11a,ramos11b,tadhunter11}. Given that hydrodynamic simulations of major, gas-rich mergers show that, as well as triggering AGN, such interactions are capable of triggering powerful starburst activity (\citealp{mihos96, dimatteo07, cox08, johansson09}), the study of the level of recent star formation activity (hereafter RSFA)\footnote{We define ``recent star formation activity'' to include all star formation activity that has occurred within  2~Gyr of the observation epoch, encompassing contemporaneous
starbursts, continuous star formation and post-starburst stellar populations (e.g. post-starburst activity as detected in spectral synthesis modelling of optical spectra: Tadhunter et al. 2005, Holt et al. 2007, Tadhunter et al. 2011).}  in radio galaxies provides key information about the nature of the triggering events, as well as the timing of the AGN relative to any merger-induced starburst.

The fact that radio-loud AGN are overwhelmingly hosted by early-type galaxies allows particularly clean searches to be made for the signs of RSFA. Perhaps surprisingly, given the evidence that many of them have been involved in galaxy interactions (\citealp{heckman86,ramos11a,ramos11b}), only a minority of radio galaxies are found to show spectroscopic evidence for RSFA at optical wavelengths \citep{tadhunter02,tadhunter11}. However, it is possible that a substantial percentage of any star formation activity is obscured at short wavelengths by circum-nuclear dust. Moreover, in some cases  the optical continuum features that are characteristic of RSFA may be masked by AGN-related continuum emission components such as direct or scattered AGN light, and nebular continuum \citep{tadhunter02}. Clearly, it is important to explore star formation diagnostics that are less sensitive to dust extinction and AGN-related continuum emission. Therefore we are undertaking a programme to investigate the degree of RSFA in radio galaxies using mid- to far-IR (MFIR) diagnostics. 
 
As a first step we used the Multiband Imaging Photometer for Spitzer (MIPS: \citealp{rieke04}) to make deep MFIR photometry measurements of the complete 2Jy sample of 46 radio-loud AGN (0.05 $< $z $<$ 0.7)  (Program 20233: PI Tadhunter;  \citealp{tadhunter07};  \citealp{dicken08,dicken09} hereafter T07, D08, D09) as well as a sample of 19 nearby 3CRR radio-loud AGN from the Spitzer archive (\citealp{dicken10}, hereafter D10). 
The results show that [OIII] optical emission line luminosity ($L_{[\rm{OIII}]}$) is strongly correlated with both the mid- (24\moo) and far-IR (70\moo) luminosities ($L_{24\mu m}$ and $L_{70\mu m}$ respectively). Since the $[\rm{OIII}]\lambda$5007 emission from the NLR provides a good indication of the intrinsic power of the illuminating AGN (e.g. \citealp{rawlings91}; \citealp{tadhunter98}; \citealp{simpson98}; \citealp{lamassa10}; and discussion in D09), the correlations between MFIR luminosity and [OIII] optical emission line luminosity provide strong empirical evidence to support AGN illumination as the dominant heating mechanism of the thermal MFIR emitting dust.  We also used energetic arguments to demonstrate that, while much of the mid-IR emission is likely to be radiated by the warm dust in the torus, the NLR clouds are a plausible location for the cooler, far-IR emitting dust (D09). 

In addition, we found evidence for enhanced far-IR emission in the minority of radio galaxies that show evidence for RSFA activity at  optical wavelengths. Our interpretation was that, while AGN illumination is the primary heating mechanism for both the warm (mid-IR emitting, 24\moo) and cool (far-IR emitting, 70\moo) dust in most powerful radio-loud AGN, heating by starbursts acts to substantially boost the 70\moo\ luminosity in the 20-30$\%$ of objects in the 2Jy sample with optical evidence for RSFA. 

Unfortunately, apart from the most extreme objects with the highest degrees of star formation activity, the far-IR excess does not necessarily provide an accurate indication of the level of star formation activity in individual radio galaxies. This is because the degree of contamination of the far-IR continuum by emission from the NLR and circum-nuclear torus may vary substantially from object to object. An alternative is to use mid-IR spectroscopy to detect the strong Polycyclic Aromatic Hydrocarbon (PAH) emission features at 6.2, 7.7, 8.6 and 11.3\moo. These well-defined emission bands are considered to provide an unambiguous signature of RSFA \citep{diamond10}, and are much less susceptible to dust extinction effects than optical continuum studies. 

Therefore we have undertaken a campaign to observe the complete 2Jy sample of 46 objects ($0.05 < z < 0.7$), previously observed with Spitzer/MIPS, using the Spitzer Infrared Spectrograph (IRS: \citealp{houck04};  program 50588: PI Tadhunter ). In a similar manner to the work presented in D10, we also examine IRS spectra for a complete sample of 19 3CRR FRII radio galaxies that have, on average, lower redshifts and radio powers than the 2Jy sample.

We aim to investigate the degree to which the presence and strength of the PAH emission features correlates with other diagnostics for identifying RSFA. In particular: is the detection of PAH features, and therefore RSFA by association, confined to the objects in which we already find evidence for such activity at optical wavelengths? Or is there a substantial population of radio galaxies in which the star formation activity is hidden by dust and/or masked by AGN-related continuum components at optical wavelengths?
 
In this, the first paper related to our IRS program of spectral observations of powerful radio galaxies, we present an atlas of Spitzer IRS mid-IR spectra for the 2Jy and 3CRR samples, and discuss the results from the PAH emission analysis. In a second paper we will discuss the silicate features and fine structure lines detected in the spectra in the context of the unified schemes for AGN. We assume a cosmology with $H_{o}=71$km s$^{-1} $Mpc$^{-1}, \Omega_{m}=0.27$ and $\Omega_{\lambda}=0.73$ throughout this work.

\section{Sample Selection}
\label{sec:sample}

The primary sample consists of all 46 powerful radio galaxies and steep-spectrum quasars ($ F_{\nu} \propto \nu^{-\alpha},\alpha^{4.8}_{2.7} > 0.5 $) selected from the 2Jy sample of \citet{wall85} with redshifts 0.05 $<$ z $<$ 0.7, flux densities $S_{2.7\rm{GHz}}>$ 2~Jy and declinations $\delta<10^o$; the sample is complete based on these criterion. 
This sample is identical to that presented in \citet{tadhunter93} and \citet{morganti93} except that the redshift and steep spectrum selection criteria have been applied, and the object PKS 0347$+$05, which has since proved to fulfill the same selection criteria \citep{diserego94}, has been added. The spectral index cut has been set to ensure that all the sources in the sample are dominated by steep spectrum lobe emission, while the lower redshift limit has been set to ensure that these galaxies are genuinely powerful sources. In the following we will refer to this sample as the 2Jy sample. 

The mid-IR spectra presented here complement a wealth of data that has been obtained for the 2Jy sample over the last two decades. These include: deep optical spectroscopic observations  (\citealp{tadhunter93, tadhunter98, tadhunter02}; \citealp{wills02}; \citealp{holt07}); extensive observations at radio wavelengths (\citealp{morganti93, morganti97, morganti99}; D08); complete deep optical imaging from Gemini \citep{ramos11a,ramos11b}; deep Spitzer/MIPS mid- to far-IR photometric observations (D08; detection rates 100\% at 24\moo\ and 90\% at 70\moo). In addition,  85\% of the sample have recently been observed with Chandra and/or XMM at X-ray wavelengths, including all objects with redshifts z $<$ 0.2; and 78\% of the sample have deep 2.2 micron (K-band) near infrared imaging \citep{inskip10}. The full sample of 46 objects includes a mixture of broad-line radio galaxies and radio-loud quasars (BLRG/Q: 35\%), narrow-line
radio galaxies (NLRG: 43\%), and weak-line radio galaxies\footnote{WLRGs are defined as
having EW([OIII]) $< $10~\AA, \citep{tadhunter98}.} (WLRG:
22\%). In terms of radio morphological classification, the sample
comprises 72\% FRII sources, 13\% FRI sources, and 15\% compact steep
spectrum (CSS)/gigahertz peak spectrum (GPS) objects. The
sample is presented in Table \ref{tbl-1}\footnote{Additional information on these objects and the more extended 2Jy sample can be found at http://2jy.extragalactic.info}.

We also present IRS spectra for a complete sub-sample of 19 3CRR radio-loud AGN selected from the sample of  \citet{laing83} (see Table \ref{tbl-2}). We have limited this sample to 3CRR objects with FRII radio morphologies and redshifts z$\leq$0.1, leading to a sample which is complete in both Spitzer/MIPS detections (100\% at 24\moo\ and 89\% at 70\moo) and [OIII] $\lambda5007$ emission line flux measurements (100\%). The full sample of 19 objects also includes a mixture of BLRG/Q (16\%), NLRG (58\%), WLRG (26\%). Because the 3CRR objects have lower radio powers and redshifts on average than most of the 2Jy sample, they help to fill in the lower luminosity ends of the MFIR vs [OIII] correlations (D10). In the following discussion we will refer to this sample as the 3CRR sample. Note that, although two objects in the sample (3C277.3, 3C293) have uncertain radio morphological classifications, and cannot  be confidently characterized as either FRI or FRII types, they are included here for completeness.
Also, the compact steep spectrum radio source 3C305 is included because it presents a miniature FRII radio morphology, with distinct hotspots at the ends of its jets, even if its radio structure as a whole is highly distorted because of strong jet-cloud interactions \citep{heckman82}. The Spitzer/MIPS flux data and associated errors are presented in D10. The [OIII] fluxes were obtained from published deep optical spectra at both high and low resolution taken using Dolores on the TNG \citep{buttiglione09}, except for DA240, 4C73.08, 3C321 and 3C445 (see D10 for details). Note that two objects overlap between the 3CRR and 2Jy samples (3C403, 3C445). 

\begin{deluxetable}{c@{\hspace{-1mm}}c@{\hspace{-1mm}}c@{\hspace{-1mm}}c@{\hspace{-3mm}}c@{\hspace{-3mm}}c@{\hspace{-3mm}}c@{\hspace{-3mm}}c@{\hspace{-3mm}}c@{\hspace{-3mm}}c@{\hspace{-3mm}}c@{\hspace{-3mm}}c@{\hspace{-3mm}}c@{\hspace{-3mm}}c}
\tabletypesize{\scriptsize}
\tablecaption{The 2Jy Sample - Spitzer Observational Data \label{tbl-1}}
\tablewidth{0pt}
\tablehead{
\colhead{Name}{\hspace{-1mm}} &\colhead{Other name}{\hspace{-1mm}} &\colhead{z}{\hspace{-1mm}} &\colhead{Mode}{\hspace{-3mm}} &\colhead{Obs. date}{\hspace{-3mm}} &
\colhead{$t_{int}$(SL1)}{\hspace{-3mm}} & \colhead{cycles}{\hspace{-3mm}} & \colhead{$t_{int}$(SL2)}{\hspace{-3mm}}  & \colhead{cycles}{\hspace{-3mm}} & \colhead{$t_{int}$(LL1)}{\hspace{-3mm}} & \colhead{cycles}{\hspace{-3mm}} & \colhead{$t_{int}$(LL2)}{\hspace{-3mm}} & \colhead{cycles}& \colhead{xSL}
}
\startdata
0023$-$26	&	\phantom{a}		&	0.322	&	S	&	Jan-09	&	60.0	&	8	&	60	&	8	&	120	&	4	&	120	&	4	&	-	\\
0034$-$01	&	\phantom{a}	3C15	&	0.073	&	S	&	Jan-06	&	240	&	2	&	240	&	3	&	120	&	3	&	120	&	2	&	0.8	\\
0035$-$02	&	\phantom{a}	3C17	&	0.220	&	S	&	Jan-06	&	240	&	2	&	240	&	3	&	120	&	3	&	120	&	2	&	-	\\
0038$+$09	&	\phantom{a}	3C18	&	0.188	&	S	&	Jan-09	&	14	&	16	&	14	&	16	&	30	&	4	&	30	&	4	&	-	\\
0039$-$44	&	\phantom{a}		&	0.346	&	S	&	Dec-08	&	60	&	4	&	60	&	4	&	30	&	4	&	30	&	4	&	1.1	\\
0043$-$42	&	\phantom{a}		&	0.116	&	S	&	Dec-08	&	60	&	4	&	60	&	4	&	120	&	2	&	120	&	2	&	-	\\
0105$-$16	&	\phantom{a}	3C32	&	0.400	&	S	&	Jan-09	&	60	&	4	&	60	&	4	&	120	&	2	&	120	&	2	&	-	\\
0117$-$15	&	\phantom{a}	3C38	&	0.565	&	S	&	Jan-09	&	60	&	8	&	60	&	8	&	120	&	4	&	120	&	4	&	-	\\
0213$-$13	&	\phantom{a}	3C62	&	0.147	&	S	&	Jan-09	&	60	&	2	&	60	&	2	&	30	&	4	&	30	&	4	&	1.1	\\
0235$-$19	&	\phantom{a}	OD-159	&	0.620	&	S	&	Oct-08	&	60	&	4	&	60	&	4	&	120	&	2	&	120	&	2	&	-	\\
0252$-$71	&	\phantom{a}		&	0.566	&	S	&	Oct-08	&	240	&	2	&	240	&	2	&	120	&	4	&	120	&	4	&	-	\\
0347$+$05	&	\phantom{a}		&	0.339	&	S	&	Oct-08	&	60	&	8	&	60	&	8	&	120	&	4	&	120	&	4	&	-	\\
0349$-$27	&	\phantom{a}		&	0.066	&	S	&	Oct-08	&	60	&	4	&	60	&	4	&	120	&	2	&	120	&	2	&	-	\\
0404$+$03	&	\phantom{a}	3C105	&	0.089	&	M	&	Sep-05	&	14	&	1	&	14	&	1	&	14	&	1	&	14	&	1	&	-	\\
0409$-$75	&	\phantom{a}		&	0.693	&	S	&	Dec-08	&	240	&	2	&	240	&	2	&	120	&	4	&	120	&	4	&	-	\\
0442$-$28	&	\phantom{a}		&	0.147	&	S	&	Nov-08	&	60	&	4	&	60	&	4	&	30	&	4	&	30	&	4	&	-	\\
0620$-$52	&	\phantom{a}		&	0.051	&	S	&	Dec-08	&	240	&	2	&	240	&	2	&	120	&	3	&	120	&	3	&	1.2	\\
0625$-$35	&	\phantom{a}	OH-342	&	0.055	&	S	&	Dec-08	&	60	&	2	&	60	&	2	&	30	&	4	&	30	&	4	&	1.1	\\
0625$-$53	&	\phantom{a}		&	0.054	&	S	&	Dec-08	&	240	&	2	&	240	&	2	&	120	&	4	&	120	&	4	&	-	\\
0806$-$10	&	\phantom{a}	3C195	&	0.110	&	S	&	Dec-08	&	14	&	4	&	6	&	2	&	6	&	2	&	6	&	2	&	1.1	\\
0859$-$25	&	\phantom{a}		&	0.305	&	S	&	Jan-09	&	60	&	4	&	60	&	4	&	120	&	2	&	120	&	2	&	1.3	\\
0915$-$11	&	\phantom{a}	Hydra A	&	0.054	&	S	&	Dec-05	&	60	&	9	&	60	&	11	&	30	&	4	&	30	&	4	&	3.0	\\
0945$+$07	&	\phantom{a}	3C227	&	0.086	&	M	&	May-06	&	14	&	1	&	14	&	1	&	14	&	1	&	14	&	1	&	1.2	\\
1136$-$13	&	\phantom{a}		&	0.554	&	S	&	Jul-08	&	60	&	4	&	60	&	4	&	120	&	2	&	120	&	2	&	-	\\
1151$-$34	&	\phantom{a}		&	0.258	&	S	&	Feb-09	&	60	&	2	&	60	&	2	&	30	&	4	&	30	&	4	&		\\
1306$-$09	&	\phantom{a}		&	0.464	&	S	&	Aug-08	&	14	&	32	&	14	&	32	&	120	&	3	&	120	&	3	&	-	\\
1355$-$41	&	\phantom{a}		&	0.313	&	S	&	Mar-09	&	60	&	2	&	60	&	2	&	30	&	4	&	30	&	4	&	1.1	\\
1547$-$79	&	\phantom{a}		&	0.483	&	S	&	Oct-08	&	60	&	4	&	60	&	4	&	120	&	2	&	120	&	2	&	1.1	\\
1559$+$02	&	\phantom{a}	3C327	&	0.104	&	S	&	Oct-08	&	14	&	4	&	6	&	2	&	6	&	2	&	6	&	2	&	1.1	\\
1602$+$01	&	\phantom{a}	3C327.1	&	0.462	&	S	&	Mar-06	&	240	&	2	&	240	&	3	&	120	&	3	&	120	&	2	&	-	\\
1648$+$05	&	\phantom{a}	Herc A	&	0.154	&	S	&	-	&	-	&	-	&	-	&	-	&	-	&	-	&	-	&	-	&	-	\\
1733$-$56	&	\phantom{a}		&	0.098	&	S	&	Oct-08	&	60	&	2	&	60	&	2	&	30	&	2	&	30	&	2	&	-	\\
1814$-$63	&	\phantom{a}		&	0.063	&	S	&	Nov-08	&	60	&	2	&	60	&	2	&	30	&	2	&	20	&	2	&	1.1	\\
1839$-$48	&	\phantom{a}		&	0.112	&	S	&	Nov-08	&	60	&	6	&	60	&	6	&	120	&	4	&	120	&	4	&	-	\\
1932$-$46	&	\phantom{a}		&	0.231	&	S	&	Nov-08	&	60	&	8	&	60	&	8	&	120	&	4	&	120	&	4	&	0.7	\\
1934$-$63	&	\phantom{a}		&	0.183	&	S	&	Nov-08	&	60	&	4	&	60	&	4	&	30	&	4	&	30	&	4	&	-	\\
1938$-$15	&	\phantom{a}		&	0.452	&	S	&	Nov-08	&	14	&	16	&	14	&	16	&	120	&	2	&	120	&	2	&	1.8	\\
1949$+$02	&	\phantom{a}	3C403	&	0.059	&	M	&	Oct-05	&	14	&	1	&	14	&	1	&	14	&	1	&	14	&	1	&	-	\\
1954$-$55	&	\phantom{a}		&	0.060	&	S	&	Nov-08	&	60	&	6	&	60	&	6	&	120	&	4	&	120	&	4	&	-	\\
2135$-$14	&	\phantom{a}		&	0.200	&	S	&	-	&	-	&	-	&	-	&	-	&	-	&	-	&	-	&	-	&	-	\\
2135$-$20	&	\phantom{a}	OX-258	&	0.635	&	S	&	Nov-08	&	14	&	28	&	14	&	28	&	120	&	4	&	120	&	4	&	-	\\
2211$-$17	&	\phantom{a}	3C444	&	0.153	&	S	&	-	&	-	&	-	&	-	&	-	&	-	&	-	&	-	&	-	&	-	\\
2221$-$02	&	\phantom{a}	3C445	&	0.057	&	S	&	Jun-04	&	30	&	4	&	60	&	2	&	14	&	2	&	14	&	2	&	1.2	\\
2250$-$41	&	\phantom{a}		&	0.310	&	S	&	Nov-08	&	60	&	4	&	60	&	4	&	120	&	3	&	120	&	3	&	-	\\
2314$+$03	&	\phantom{a}	3C459	&	0.220	&	S	&	Jan-09	&	14	&	16	&	14	&	16	&	30	&	2	&	30	&	2	&	-	\\
2356$-$61	&	\phantom{a}		&	0.096	&	S	&	Nov-08	&	60	&	4	&	60	&	4	&	30	&	4	&	30	&	4	&	-	\\
\enddata

\tablecomments{Table presenting the basic parameters for the 2Jy sample as well as details of the Spitzer/IRS observations. Column 4 presents the IRS observing mode: S -- Staring mode; M -- Mapping mode. Integration time in seconds and cycles for the {\it IRS} observations are presented in columns 6-13. SL1: 5.2$-$8.7\moo; SL2: 7.4$-$14.5\moo; LL1: 14$-$21.3\moo; LL2:19.5$-$38. Column 14 presents the multiplication factor applied to the SL spectrum in order to match the LL spectrum where applicable.  }  
\end{deluxetable}

\begin{deluxetable}{c@{\hspace{-1mm}}c@{\hspace{-1mm}}c@{\hspace{-3mm}}c@{\hspace{-3mm}}c@{\hspace{-3mm}}c@{\hspace{-3mm}}c@{\hspace{-3mm}}c@{\hspace{-3mm}}c@{\hspace{-3mm}}c@{\hspace{-3mm}}c@{\hspace{-3mm}}c@{\hspace{-3mm}}c}
\tabletypesize{\scriptsize}
\tablecaption{The 3CRR Sample - Spitzer Observational Data \label{tbl-2}}
\tablewidth{0pt}
\tablehead{
\colhead{Name}{\hspace{-1mm}} &\colhead{z}{\hspace{-1mm}} &\colhead{Mode}{\hspace{-3mm}} &\colhead{Obs. date}{\hspace{-3mm}} &
\colhead{$t_{int}$(SL1)}{\hspace{-3mm}} & \colhead{cycles}{\hspace{-3mm}} & \colhead{$t_{int}$(SL2)}{\hspace{-3mm}}  & \colhead{cycles}{\hspace{-3mm}} & \colhead{$t_{int}$(LL1)}{\hspace{-3mm}} & \colhead{cycles}{\hspace{-3mm}} & \colhead{$t_{int}$(LL2)}{\hspace{-3mm}} & \colhead{cycles}& \colhead{xSL}
}
\startdata
3C33	&	\phantom{a}	0.060	&	S	&	Jan-05	&	60	&	2	&	60	&	2	&	120	&	2	&	120	&	1	&	1.0	\\
3C35	&	\phantom{a}	0.067	&	M	&	Jan-06	&	14	&	1	&	14	&	1	&	14	&	1	&	14	&	1	&	--	\\
3C98	&	\phantom{a}	0.030	&	M	&	Sep-06	&	14	&	1	&	14	&	1	&	14	&	1	&	14	&	1	&	--	\\
DA240	&	\phantom{a}	0.036	&	--	&	--	&	--	&	--	&	--	&	--	&	--	&	--	&	--	&	--	&	--	\\
3C192	&	\phantom{a}	0.060	&	S	&	Nov-05	&	60	&	2	&	60	&	2	&	120	&	2	&	120	&	1	&	--	\\
4C73.08	&	\phantom{a}	0.058	&	--	&	--	&	--	&	--	&	--	&	--	&	--	&	--	&	--	&	--	&	--	\\
3C236	&	\phantom{a}	0.101	&	M	&	Dec-05	&	14	&	1	&	14	&	1	&	14	&	1	&	14	&	1	&	--	\\
3C277.3	&	\phantom{a}	0.085	&	--	&	Jan-06	&	14	&	1	&	14	&	1	&	14	&	1	&	14	&	1	&	--	\\
3C285	&	\phantom{a}	0.079	&	M	&	Jan-06	&	14	&	1	&	14	&	1	&	14	&	1	&	14	&	1	&	--	\\
3C293	&	\phantom{a}	0.045	&	S	&	Jan-06	&	240	&	2	&	240	&	3	&	120	&	3	&	120	&	2	&	1.2	\\
3C305	&	\phantom{a}	0.042	&	M	&	Apr-06	&	14	&	1	&	14	&	1	&	14	&	1	&	14	&	1	&	--	\\
3C321	&	\phantom{a}	0.096	&	S	&	Feb-05	&	14	&	2	&	14	&	2	&	30	&	1	&	30	&	1	&	--	\\
3C326	&	\phantom{a}	0.090	&	S	&	Mar-05	&	60	&	2	&	60	&	2	&	120	&	2	&	120	&	1	&	--	\\
3C382	&	\phantom{a}	0.058	&	S	&	Aug-05	&	60	&	2	&	60	&	2	&	120	&	2	&	120	&	1	&	1.2	\\
3C388	&	\phantom{a}	0.092	&	S	&	Jul-05	&	60	&	2	&	60	&	2	&	120	&	2	&	120	&	1	&	1.5	\\
3C390.3	&	\phantom{a}	0.056	&	S	&	Aug-04	&	240	&	1	&	240	&	1	&	120	&	4	&	120	&	4	&	1.0	\\
3C452	&	\phantom{a}	0.081	&	S	&	Dec-04	&	60	&	2	&	60	&	2	&	120	&	2	&	120	&	1	&	--	\\
\enddata

\tablecomments{Table presenting the basic parameters for the 3CRR sample as well as details of the Spitzer/IRS observations. Note that 3C403 and 3C445 overlap between the two samples, see Table \ref{tbl-1} for observation details of these two objects.  Definitions are the same as Table \ref{tbl-1}}  
\end{deluxetable}

\section{Observations and data reduction}
\label{sec:2}

\begin{figure*}[t]
\epsscale{2.3}
\plottwo{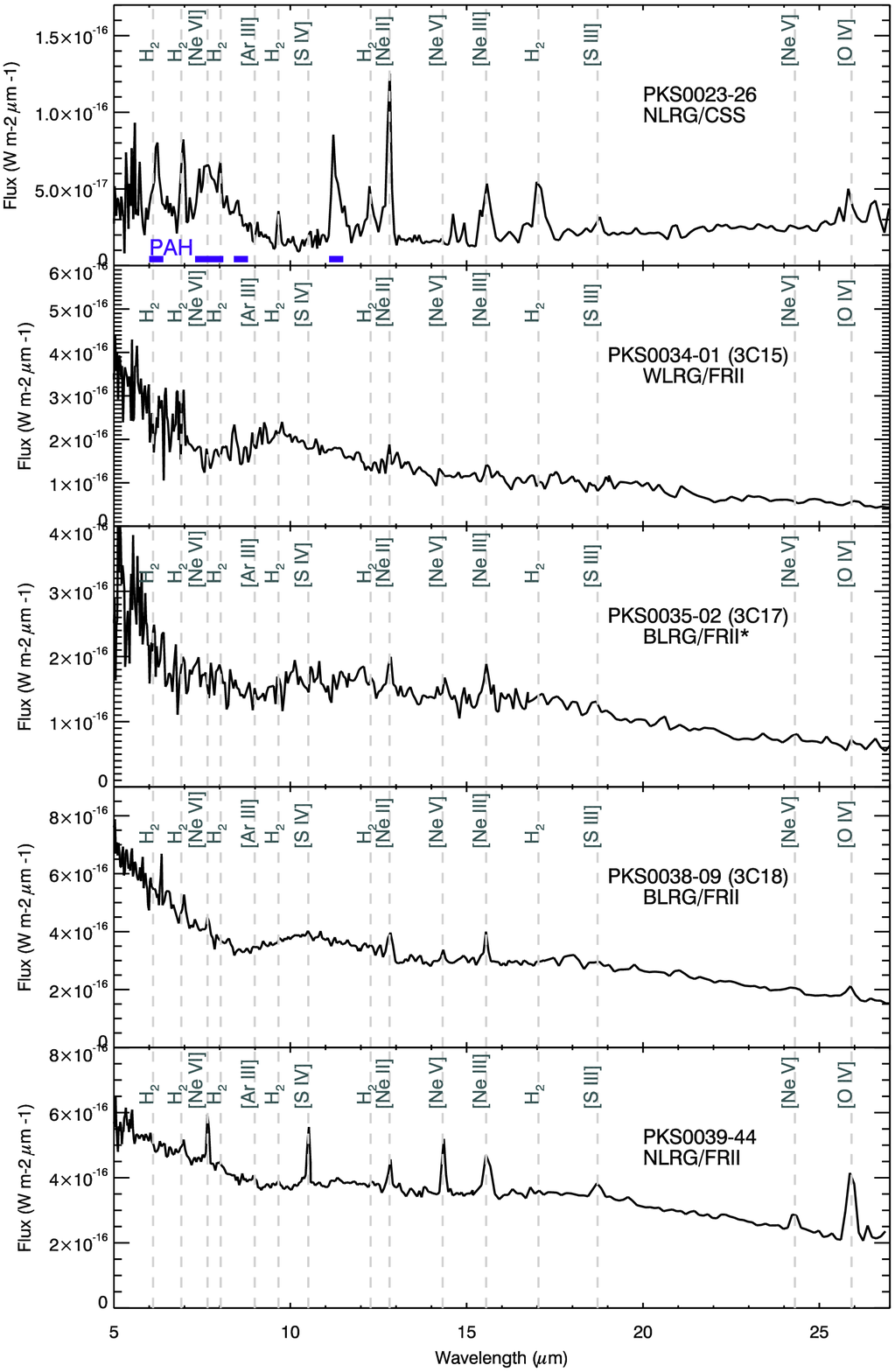}{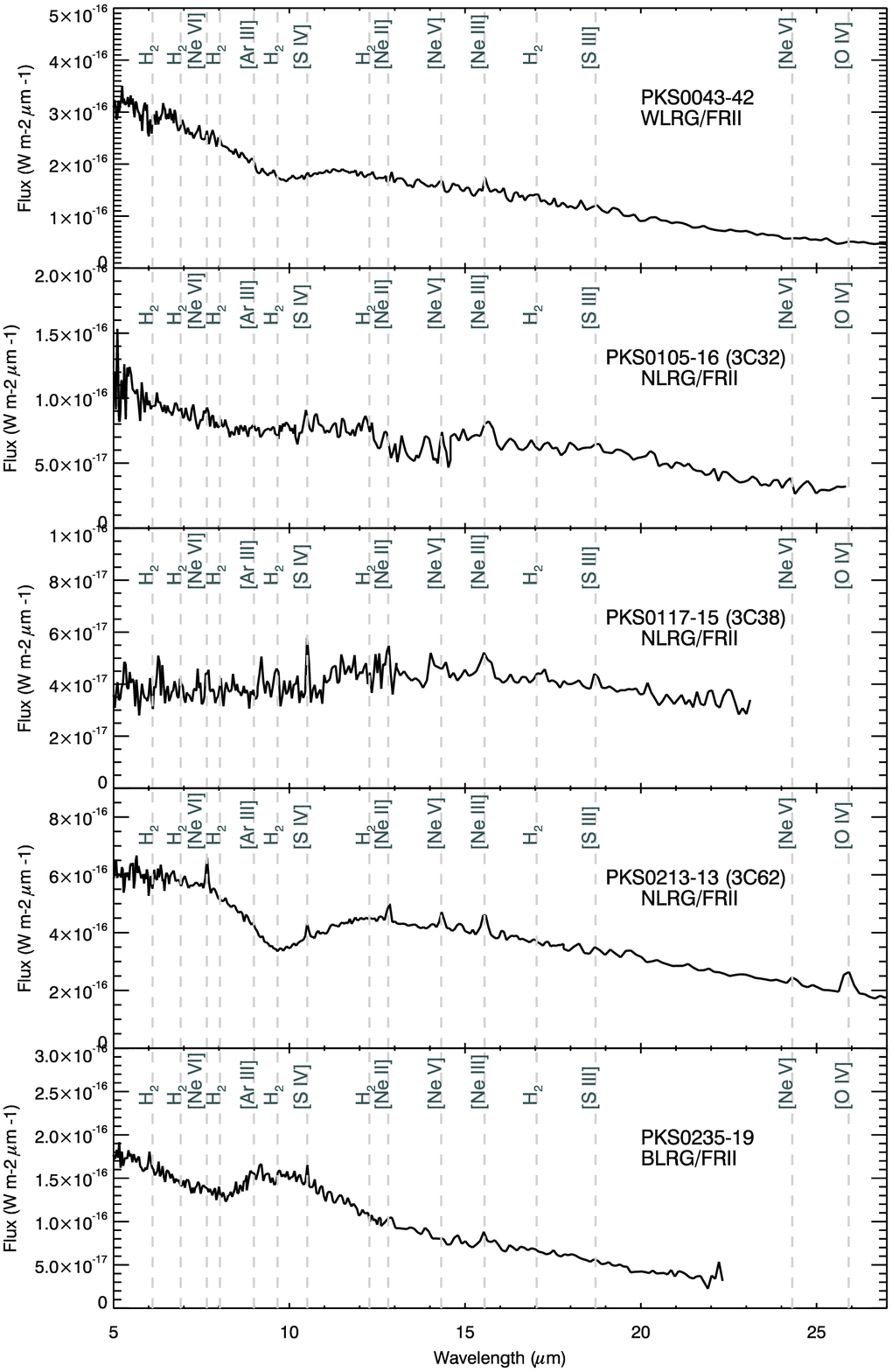}
\caption{Spitzer/IRS spectra for the 2Jy sample. Common fine structure and H$_2$ emission lines are indicated by vertical grey dashed lines. In objects with detected PAH features, the most prominent PAH features are indicated by blue line segments. Note the data for PKS0034-01 and PKS0035-02 were obtained from the Spitzer archive. Both potentially suffer from enhanced flux calibration uncertainties at short wavelengths, due to saturation of the peak-up detector (see Section \ref{sec:2} for details). \label{fig1} }
\end{figure*}

\begin{figure*}[t]
\epsscale{2.3}
\plottwo{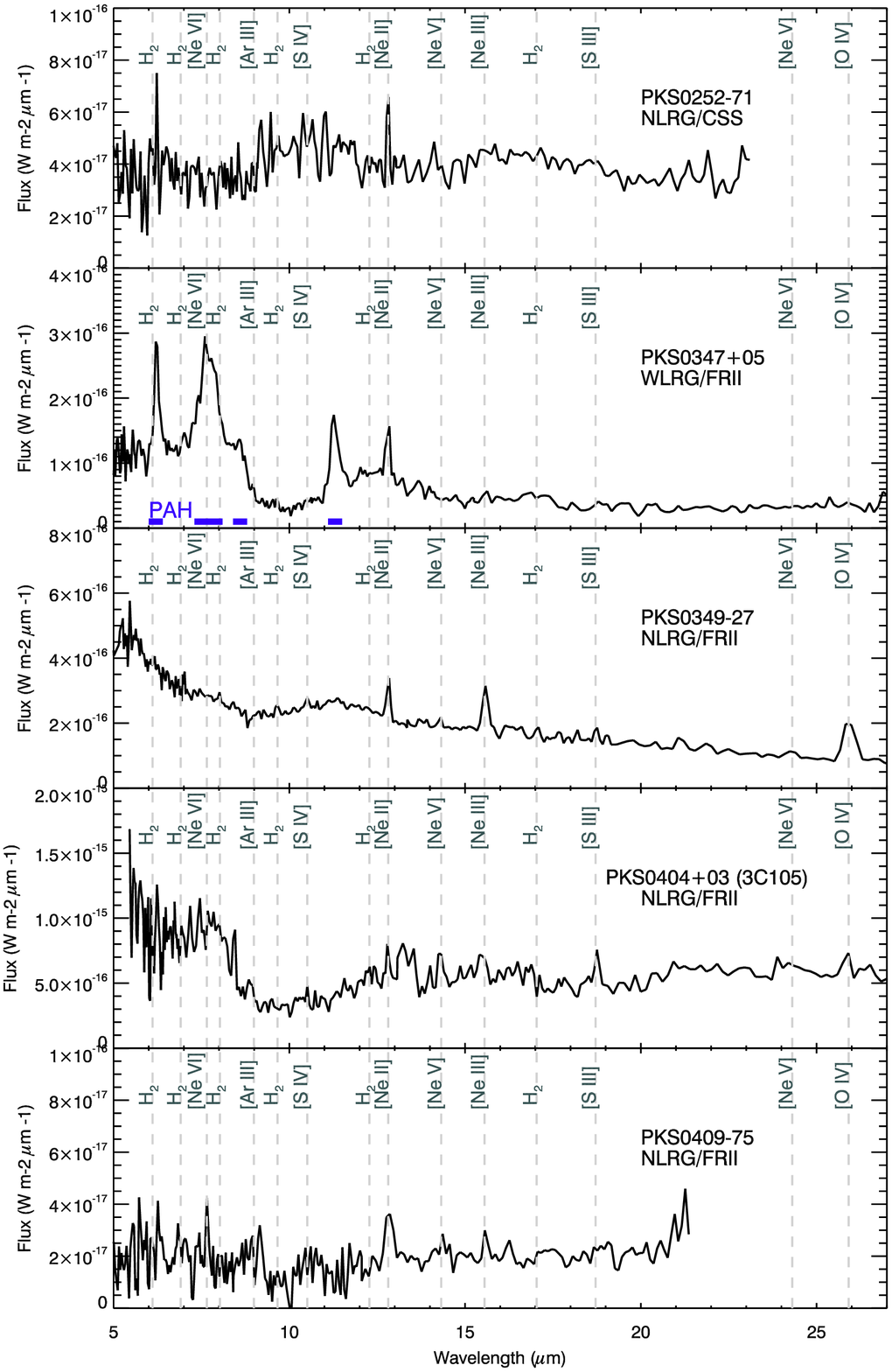}{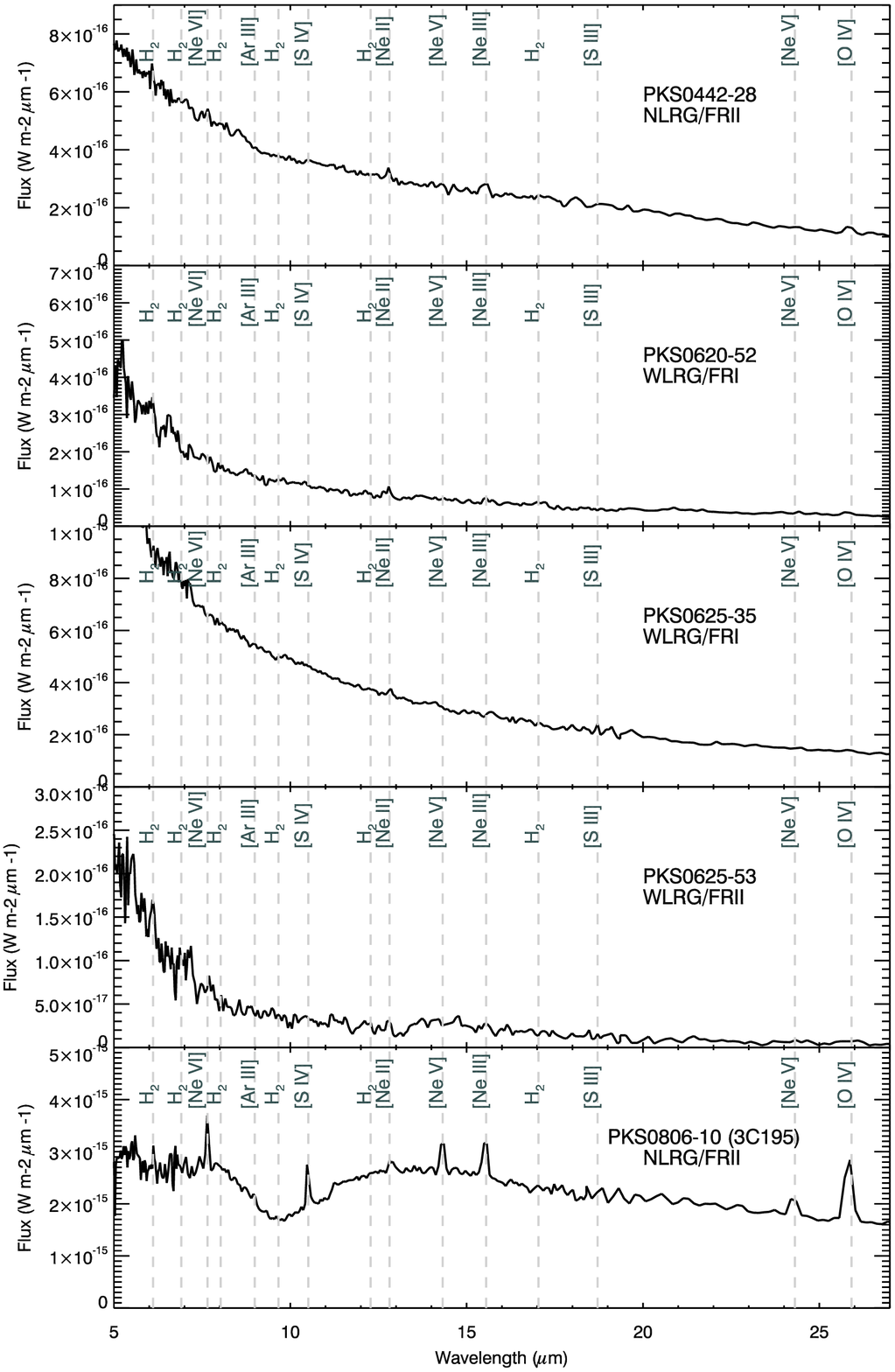}
\caption{Spitzer/IRS spectra for the 2Jy sample continued. Note the data for PKS0404+03  were taken in Mapping Mode and were obtained from the Spitzer archive. \label{fig2} }
\end{figure*}

\begin{figure*}[t]
\epsscale{2.3}
\plottwo{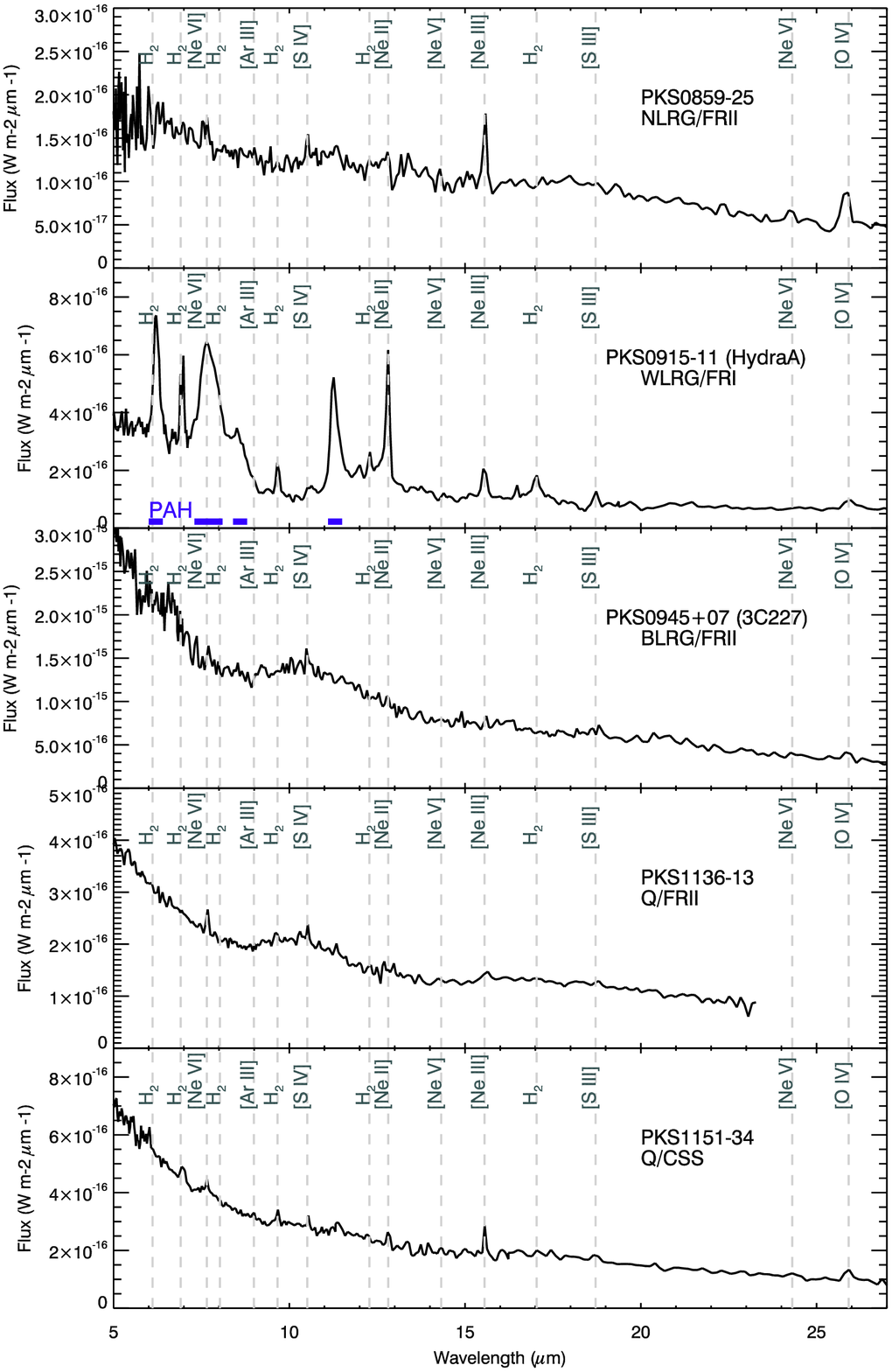}{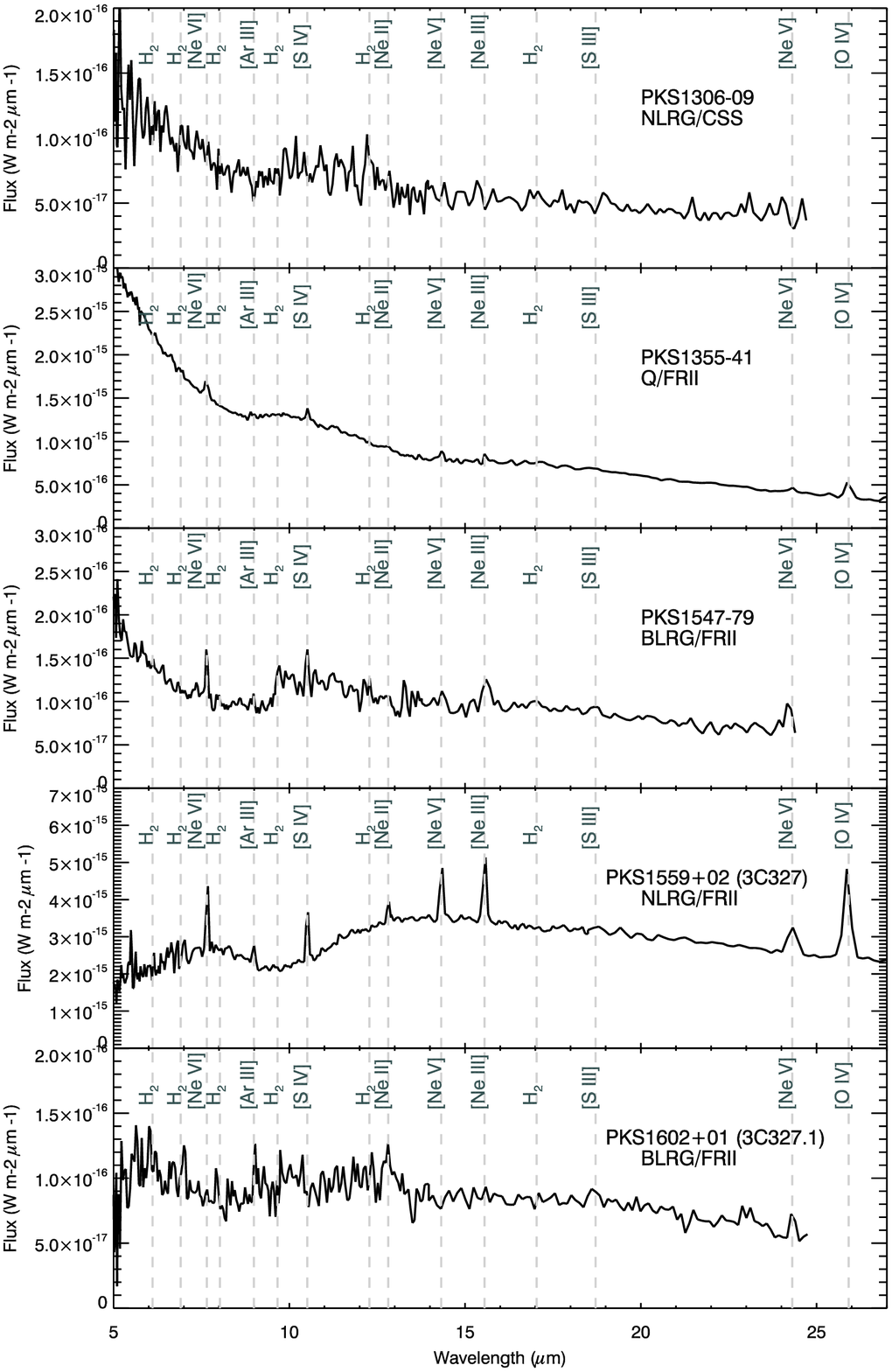}
\caption{Spitzer/IRS spectra for the 2Jy sample continued. Note the data for PKS0915-11, PKS0945+07 and PKS1602+01 were obtained from the Spitzer archive, and the data for PKS0947+07 were taken in Mapping Mode. Potentially, the spectrum of PKS1602+01 suffers from enhanced flux calibration uncertainties at short wavelengths, due to saturation of the peak-up detector.\label{fig3} }
\end{figure*}

\begin{figure*}[t]
\epsscale{2.3}
\plottwo{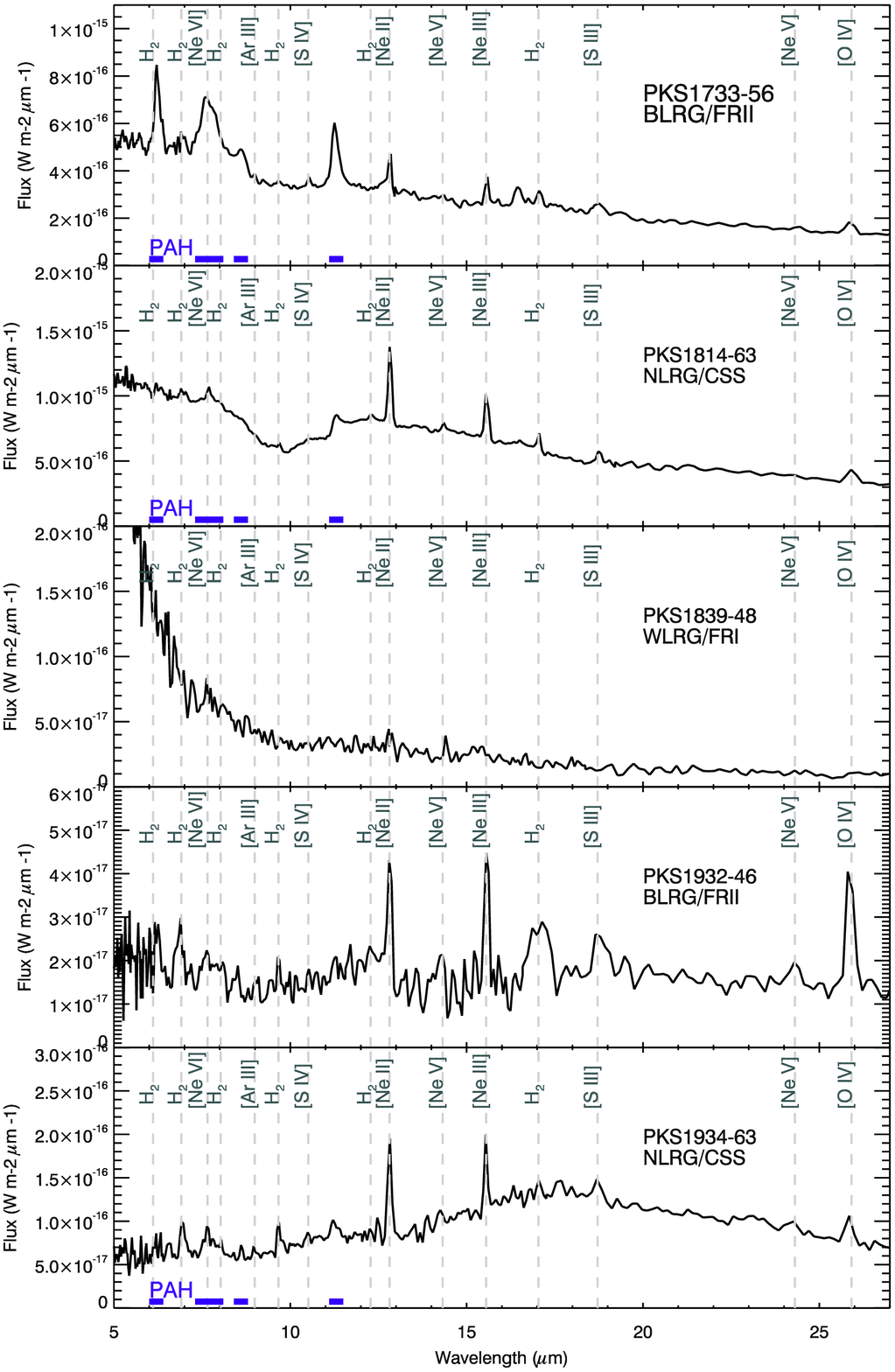}{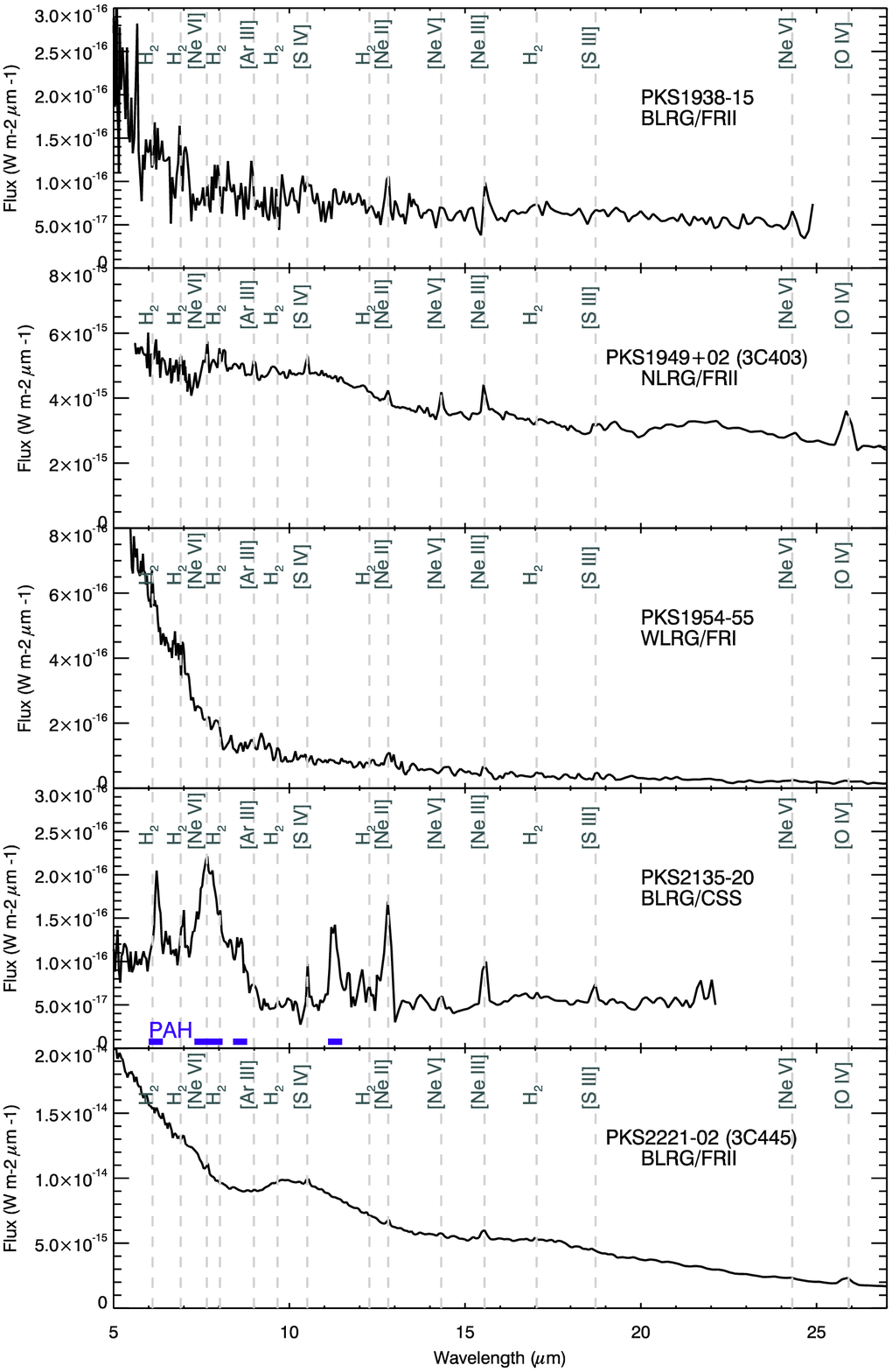}
\caption{Spitzer/IRS spectra for the 2Jy sample continued. Note that  the data for PKS1949+02 (Mapping Mode) and PKS2221-02 were obtained from the Spitzer archive. \label{fig4} }
\end{figure*}

\begin{figure*}[t]
\epsscale{2.3}
\plottwo{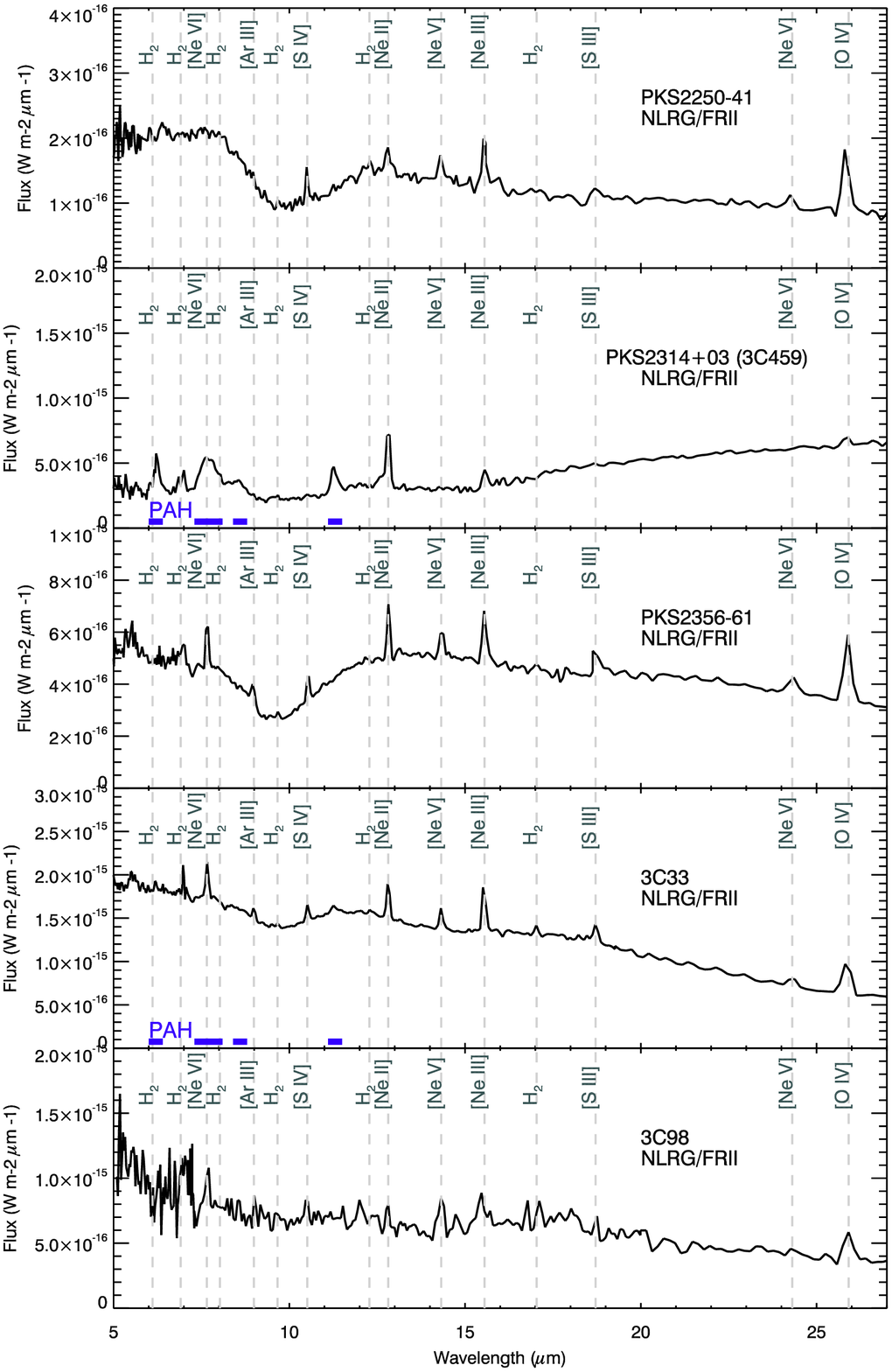}{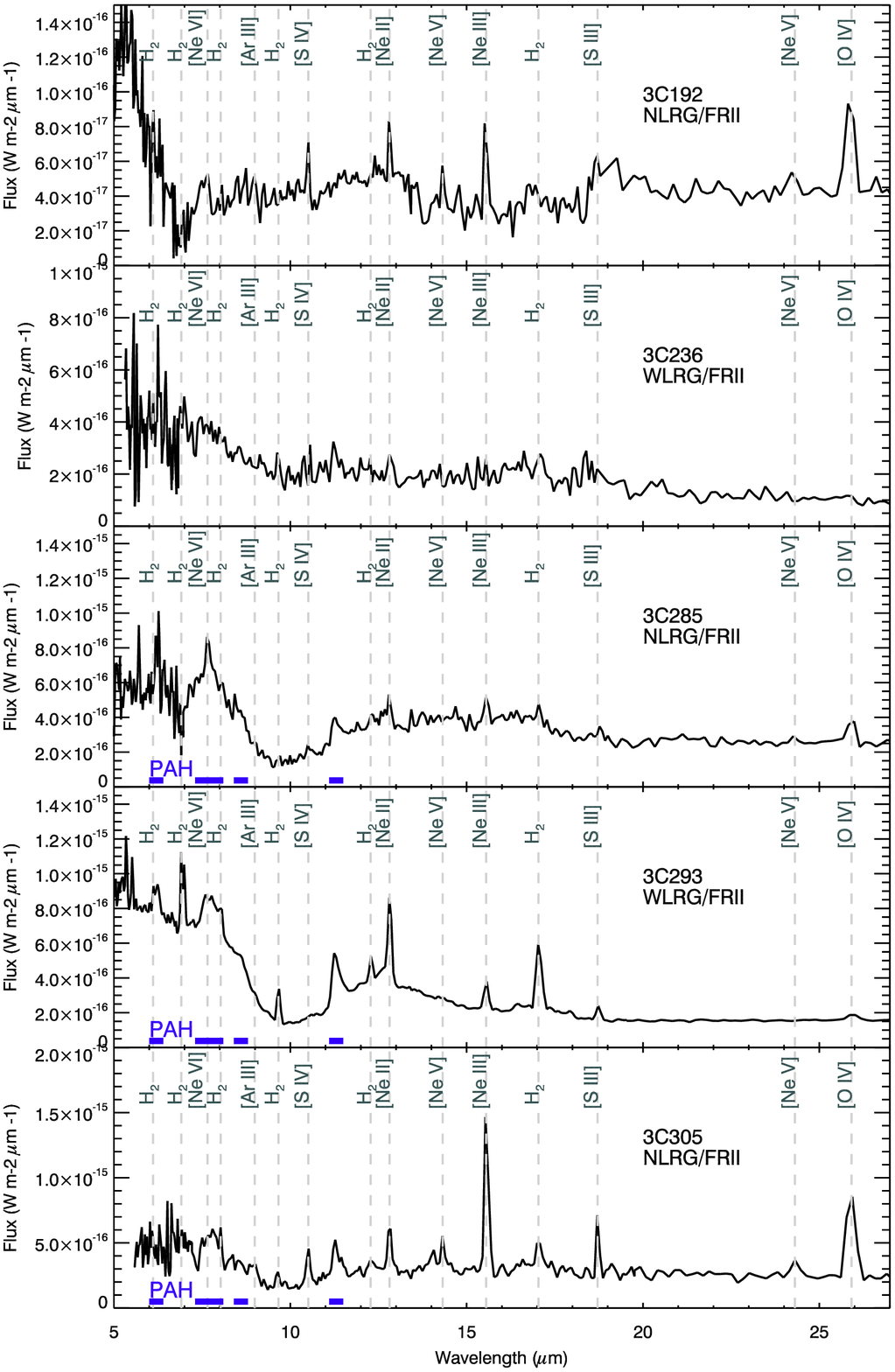}
\caption{Spitzer/IRS spectra for the 2Jy and 3CRR sample objects. All the 3CRR IRS data were obtained from the Spitzer archive. The data for 3C98, 3C236, 3C277.3, 3C285 and 3C305 were taken in Mapping Mode and were obtained from the Spitzer archive. \label{fig5} }
\end{figure*}

The mid-IR spectra of 35 out of the 46 objects in the 2Jy sample were obtained in a dedicated campaign of Spitzer IRS observations (program 50558; P.I. Tadhunter) between July 2008 and March 2009. 
These data were taken in the short low (SL) and long low (LL) resolution staring modes, covering 5.2 to 14.5\moo\ and 14.0 to 38.0\moo\ respectively. The resolution for the observations was R$\approx$60-127 and the slit widths were $3.6^{\prime\prime}$ and $10.5^{\prime\prime}$ for the SL and LL observations respectively. 
The spectra of 8 additional objects in the 2Jy sample, taken between June 2004 and May 2006, were obtained from the Spitzer archive. These data came from various campaigns under several different PIs and, consequently, have varying integration times and observing modes. Details of the observations can be found in Table \ref{tbl-1}. 

Out of the 46 objects in the 2Jy sample, 43 (93\%) were successfully detected by Spitzer/IRS. PKS2135-14 was not observed because the Spitzer observatory ran out of cryogenic coolant before the observations could be made, while no observations of PKS2211-17 were taken because its mid-IR flux level (0.5 mJy at 24$\mu$m) was deemed too low to detect in reasonable integrations with Spitzer IRS. Observations of a further object  (PKS1648+05) were available in the Spitzer archive, however, the integration times were an order of magnitude lower than those used for objects of similar mid-IR flux in our program. Therefore, this source was not detected with the IRS instrument.
The SL spectra of an additional three objects (PKS0034-01, PKS0035-02, PKS1602+01) obtained from the Spitzer archive potentially suffer from higher flux calibration uncertainties, due to the effect of saturation of part of the detector used for the short wavelength range observations. 
This problem occurs because the {\it IRS} detectors encompass not only the spectral 2D images, but also the apertures used for the target  acquisition imaging. If the ambient infrared background is high (above 25 mJy/sr) then, for long integration times, the acquisition images are likely to saturate, causing the whole detector to perform irregularly.

The majority of observations (40/43) were taken in Staring Mode, in which the target is observed in several cycles of set integration time and, for each cycle, the {\it IRS} instrument nods the object between two positions on the detector. For every cycle the nod positions were subtracted (A from B, and B from A) in order to perform background subtraction on the image, as well as to counteract intrinsic sensitivity variations across the detector. The two resulting spectra were combined to produce the final spectrum. 

Spectra for three 2Jy sources, obtained from the Spitzer archive, were observed in Mapping Mode in which the spacecraft makes a pre-determined number of pointings of set integration time mapped over a target. 
The pattern can be varied, however, these three objects were all part of the same program (program 20719; P.I. Baum) and the observational patterns were identical. All pointings had an integration time of 14 secs, with 15 and 5 pointings for the SL and LL modules respectively. 
The middle number (i.e. 8 in SL, and 3 in LL) pointing was centered on the target and, since these objects are relatively bright at MFIR wavelengths, this single central observation was used to extract the spectrum of each object. Because of the relatively short exposure times of the single exposures centered on the targets, these mapping mode observations tend to have lower S/N than the staring mode observations used for the majority of the sample. In order to subtract the background, a median image of the other pointed observations was made, and we took this median image to represent the sky spectrum, subtracting this from the central pointing observation. To reduce the possibility of removing flux from the
source in this process, we avoided using those pointed observations directly either side of the central pointing in the sky image (i.e. spectra 7 and 9 for SL,  2 and 4 for LL). Overall, the continuum level of the Mapping mode data matched well with the continuum measurement from Spitzer MIPS data measured over the 24\moo\ spectral filter response range (see below for further discussion).

All the data were downloaded in the basic calibrated data (BCD) format processed with the S18.7.0 data pipeline. These data were combined and subtracted using our own IDL code and cleaned using the SSC software IRSCLEAN MASK. The data were extracted using SMART (v.8.1.2.) program developed by the IRS Team at Cornell University (\citealp{higdon04}; \citealp{lebouteiller10}). To extract the fluxes we employed the optimal extraction function which uses a super sampled PSF and weights the extracted spectra by the S/N of each pixel \citep{lebouteiller10}, assuming the objects are unresolved point sources at these wavelengths. The extracted spectra from both nod positions were compared by eye, and further cleaning by hand was performed where necessary i.e. by averaging over obvious strong sharp features that only appeared in one nod position. These spectra were then averaged together to produce the final spectrum. Note that the IRS pipeline automatically accounts for variable slit losses with wavelength. 

17 of the 19 objects in the 3CRR sample have been observed with Spitzer/IRS. Two objects were not observed (4C73.08 and DA240) and the S/N of the data for 3C35 and 3C277.3 was too low to extract science quality data.
Therefore 15/19 objects (79\%) have good IRS data. The data were reduced in an identical manner to that described for the 2Jy sample; 5 objects were observed in Mapping Mode and the rest in Staring Mode. Details of the observations can be found in Table \ref{tbl-2}. 

For the majority of the resulting spectra (70\%) the continuum level of the SL spectrum matched well with that of the LL spectrum. This demonstrates that the extraction techniques and background subtraction method are robust for most objects, and that the sources are not spatially extended. However, for 15 objects in the 2Jy sample and 2 objects in the 3CRR sample, the SL spectrum and LL spectrum did not precisely match when we attempted to combine them together. The likely causes of this flux difference include the objects being partially extended, pointing errors, or poor calibration due to peak up saturation of the detector (discussed above). For most of these objects we applied a scaling factor to the SL spectrum of between 0.8 and 1.15\footnote{For PKS0915-11 (Hydra A), PKS1938-15, 3C305, 3C382 and 3C388 the scaling factor was higher, with values in the range 1.2 --- 3.0. In these cases the objects may not have been exactly
centered in the 3.6" SL slit because of pointing errors, or the
mid-IR flux is significantly extended outside the SL slit.}, where the median scaling factor was 1.1 (see Tables \ref{tbl-1} \& \ref{tbl-2}). We kept the flux level of the LL spectrum constant because the LL slit width is larger and therefore more likely to include the entire flux from any sources that are partially extended, or not perfectly centered. 

The spatially integrated 24\moo\ flux values obtained from our Spitzer/MIPS photometric campaign can be used to check the flux calibration of the Spitzer/IRS data. The mean flux from the IRS spectra was calculated over the response range of the 24\moo\ filter for MIPS in the observed frame. We found that the estimated IRS fluxes agree to within 15\% of the measured MIPS 24\moo\ photometric values for all the sources, with a median of the ratio of $MIPS_{24\mu m}$ to $IRS_{24\mu m}$ of 0.94.  This provides evidence that our spectra capture the majority of the mid-IR emission associated with any star formation in the host galaxies of the radio sources.

Overall, we have useable IRS spectra for 56 (89\%) of the 63 objects in the combined 2Jy and 3CRR sample.

\begin{figure}
\epsscale{1}
\plotone{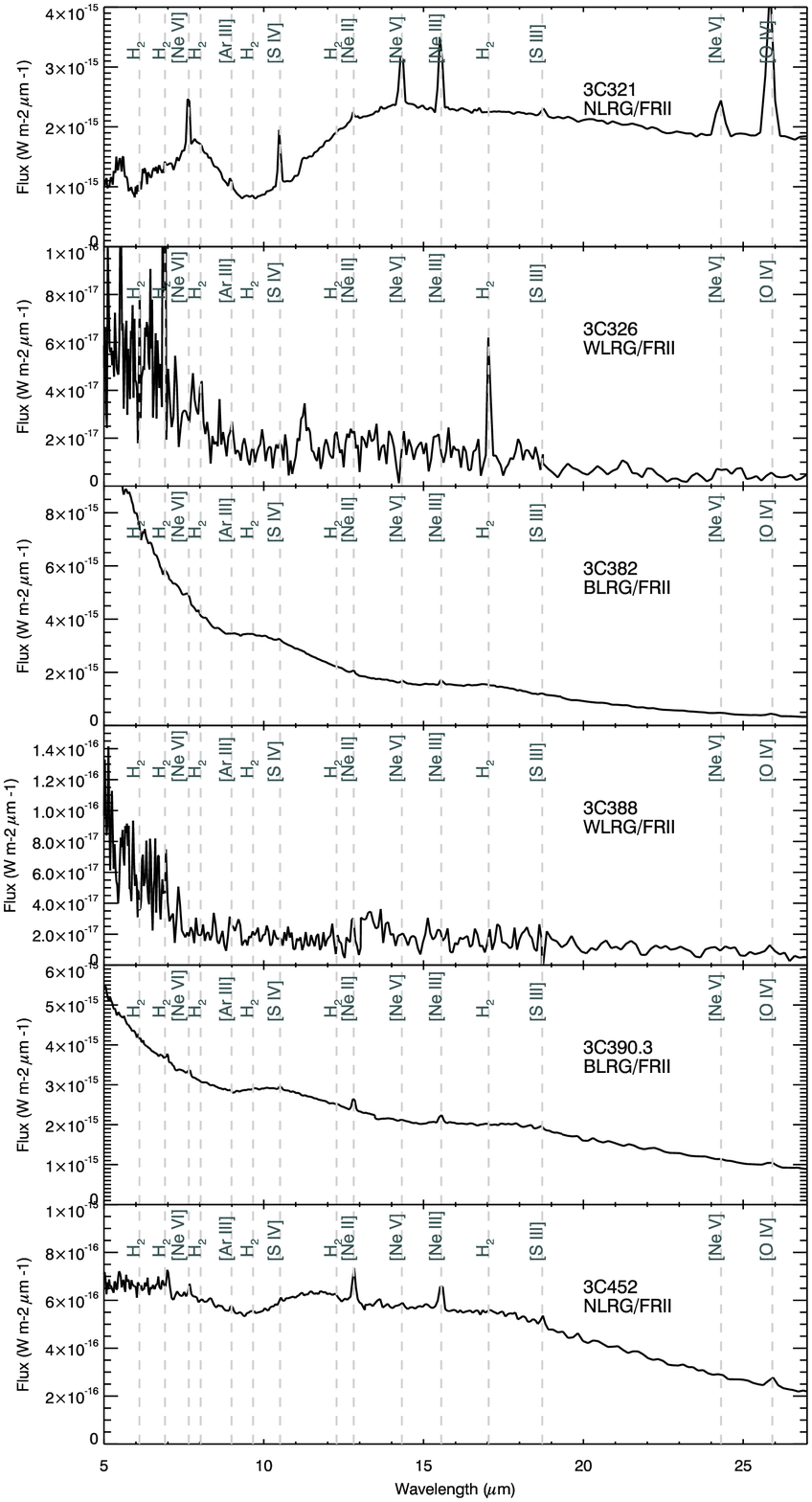}
\caption{Spitzer/IRS spectra of the 3CRR sample continued.  \label{fig6} }
\end{figure}

\section{Data Analysis}
\label{sec:3}

The fully reduced rest frame spectra for the 43 observed/detected objects in the 2Jy sample and the 13 objects from the 3CRR sample (not in common with the 2Jy sample) are presented in Figures \ref{fig1} through \ref{fig6}. The IRS  spectral range is 5-38\moo, however, at longer wavelengths the S/N is low. Therefore, we present the spectra only up to 30\moo\ in the observed frame and/or 27\moo\ in the rest frame.  

The diversity in the spectra of the radio sources shown in Figures \ref{fig1} through \ref{fig6} is striking. Many of the objects show the characteristic features of AGN at mid-IR wavelengths, including strong fine structure emission lines, as well as weak silicate emission or absorption features \citep{hao07}. We have identified some of the most prominent fine-structure lines on the spectral plots: 
[Ne VI]$\lambda$ 7.65\moo, 
[ArIII]$\lambda$8.99\moo,
[SIV]$\lambda$10.51\moo,
[NeII]$\lambda$12.81\moo,
[Ne V]$\lambda$14.32\moo,
[NeIII]$\lambda$15.56\moo,
[SIII]$\lambda$18.71\moo,
[Ne V]$\lambda$24.31\moo,
 [OIV]$\lambda$25.89\moo. We also indicate the location of $H_{2}$ lines (S(1) through S(6))\footnote{Note [ArII]$\lambda$6.99\moo\ is not indicated in the plots and may be blended with H$_2$ S(5)$\lambda$6.91\moo.}.
It is noteworthy that nearly all the spectra show the high ionization [OIV]$\lambda$25.89\moo\ ($E_{ion}$=54.9eV) emission, when not redshifted outside the observable wavelength range. This indicates a high ionization state, as expected given the presence of powerful AGN in many of the sources. Low ionization potential lines such as [NeII]$\lambda$12.81\moo\  ($E_{ion}$=21.6eV) are also detected in many spectra. The contribution of starlight from the host galaxies is significant at  short wavelengths ($\lambda < 8\mu m$) in some low redshift objects, seen as a sharp upturn in flux at the blue end of the spectra. This is particularly apparent for objects with low power AGN, for example, the WLRG PKS1839-48 and PKS1954-55 (see Figure \ref{fig4}). 
The fine structure, H$_2$, and silicate emission/absorption features will be discussed in depth in Paper II.
 
A number of PAH dust emission bands exist within the spectral range of Spitzer/IRS ($\approx$25), but at low spectral resolution they can be difficult to detect, for example the 12.7\moo\ feature which blends with the strong [NeII] line at 12.81\moo. Therefore, in this paper we focus on the three strongest PAH bands in the usable {\it IRS} wavelength range: the 6.2, 7.7, 11.3\moo\ features. The latter two PAH band are in fact blends of PAH emission (7.60 \& 7.85\moo, 11.23 \& 11.33\moo). Other bands that can make a significant contribution are the 8.6\moo\ band and a  blend of PAH emission at 17\moo\ (17.38 \& 17.87\moo), which can be strongly contaminated by $H_2$ emission. 

To fit the data and measure the PAH fluxes we used PAHFIT v1.2\footnote{PAHFIT is made available under the terms of the GNU General Public License.}, which is an IDL program developed by J.D.T Smith and B.T. Draine for studying PAH features in the mid-IR spectra of the inner regions of local star-forming galaxies \citep{smith07}. The PAHFIT model is made up of 5 components: (1) starlight continuum, represented as a blackbody with temperature T=5,000K, (2) a featureless thermal dust continuum, represented by an array of 8 different black body continuum components with temperatures from 35 to
300K, (3) pure rotational lines of $H_{2}$, (4) fine structure forbidden emission lines, and (5) PAH emission features.

We experimented with, and adapted, the PAHFIT code to check that it functioned appropriately for the samples of powerful radio-loud AGN presented in this paper. For example, we experimented with adding dust components into the model with black body temperatures 400K, 600K, 1000K  and 2000K, in order to account for the potential hotter dust continuum features in the radio galaxy samples spectra. However, PAHFIT fits the data just as well without these extra hot components, which were therefore excluded in the final fit. In addition, when fitting the 10\moo\ and 18\moo\ silicate absorption features, we found that the default PAHFIT extinction model did not fit the features well. After some experimentation it was found that the depth of the silicate absorption feature was better fitted in models assuming high ratios of 10 to 18\moo\ opacity. This implies that the 18\moo\ feature is weak for our samples of powerful radio-loud AGN. Finally, the original PAHFIT model does not include silicate emission because galaxies with such emission were not included in the original PAHFIT sample of \citet{smith07}. However, for the 2Jy sample, 10\moo\ silicate emission is clearly detected at varying strengths in 19\% of the observed/detected objects in the 2Jy and 3CRR samples. Therefore we made a further adaptation of the  PAHFIT program to fit silicate emission as well as absorption, following the method employed in \citet{gallimore10}.

\section{Results}

In this section we consider the rate of detection of PAH features in radio galaxies, and compare the various star formation diagnostics.

\subsection{PAH detection}
\label{sec:pah}

\begin{figure}
\epsscale{1}
\plotone{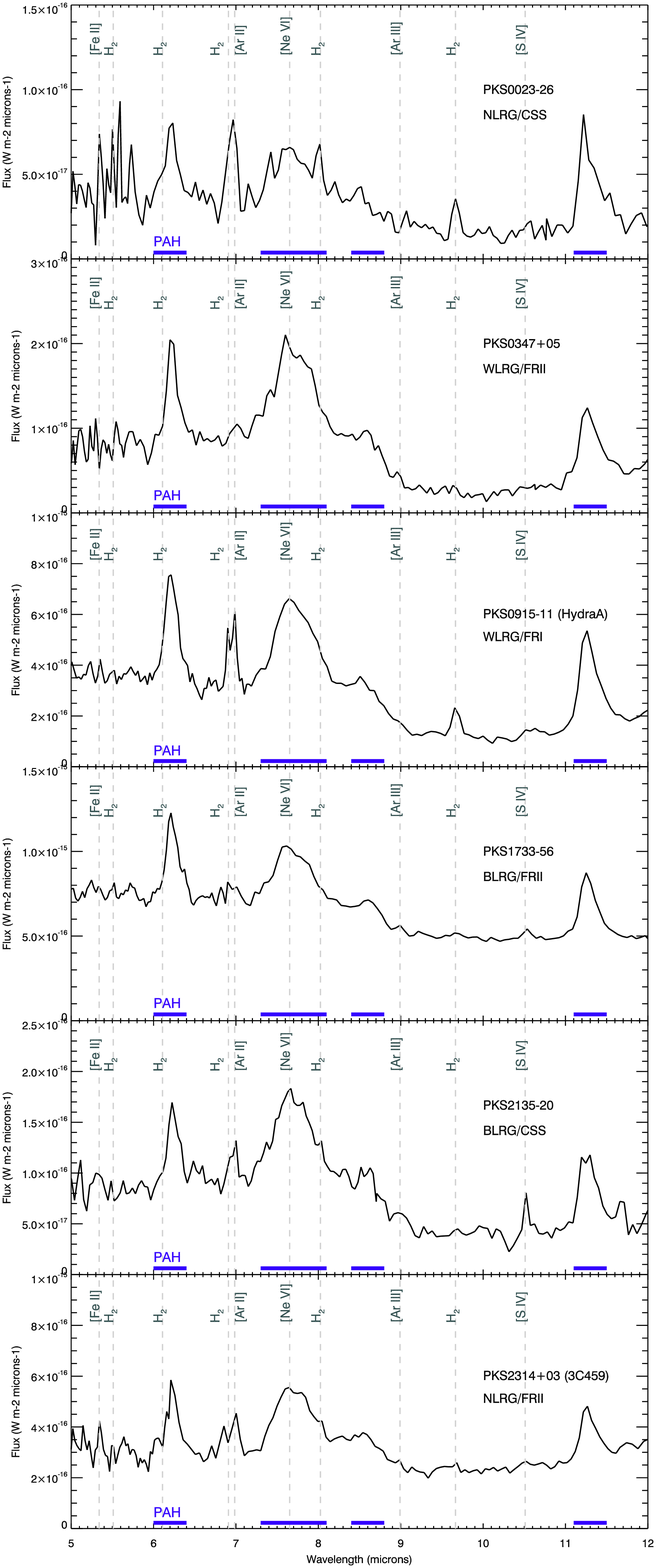}
\caption{: Magnified IRS spectra focussing on the 5-12\moo\ band of IRS to emphasize the dominant PAH bands at 6.2, 7.7 and 11.3\moo. The 6 objects in the 2Jy sample with visually identifiable PAH emission are shown. \label{fig_SLpah} }
\end{figure}

\begin{figure}
\epsscale{1}
\plotone{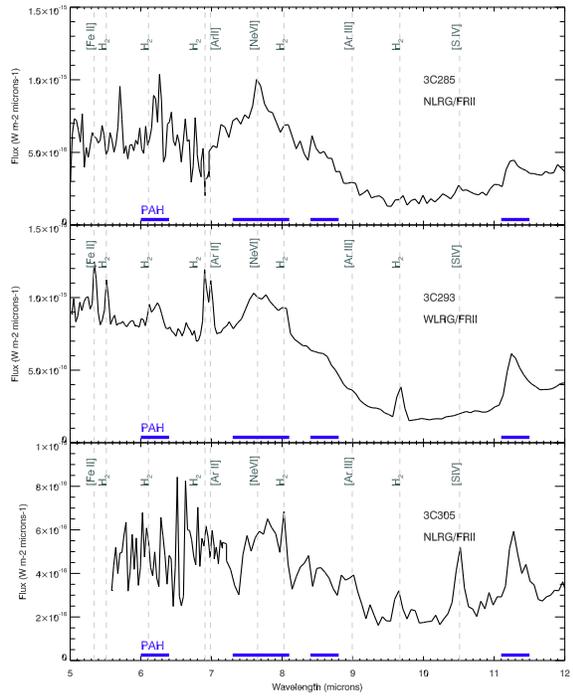}
\caption{: Magnified IRS spectra continued. Presenting the 3 objects in the 3CRR sample with visually identifiable PAH emission are shown. Note that 3C285 and 3C305 were observed in Mapping Mode and therefore suffer from lower S/N. \label{fig_SLpah2} }
\end{figure}

It is important to distinguish PAH emission features from fine-structure emission lines, in order to make secure identifications of the PAH bands. Fortunately, the PAH emission bands appear broader than the fine-structure lines, even in low resolution IRS spectra. The 11.3\moo\ PAH feature falls in a region of the spectrum without strong fine-structure emission lines, making it an ideal choice for identifying the presence of PAH emission in the radio galaxy samples.  In addition, the study of \citet{diamond10} found that the 11.3\moo\ feature is a more robust indicator of RSFA than the shorter wavelength PAH features in AGN host galaxies. Although it is more susceptible to the effects of silicate absorption and emission than shorter wavelength PAH features such as the 6.2\moo\ feature, PAHFIT accounts for the effect of silicates when fitting the 11.3\moo\ PAH emission.

For the objects with good IRS spectra in the combined 2Jy and 3CRR sample, 9 (16\%) objects have PAH features that are clearly detected at high equivalent width (PKS0023-26, PKS0347+05, PKS0915-11, PKS1733-56,  PKS2135-20, PKS2314 +03, 3C285, 3C293, 3C305). To show the PAH emission more clearly in these 9 objects we present enlargements of their spectra between 5 and 12\moo\ in Figures \ref{fig_SLpah} and \ref{fig_SLpah2} for the 2Jy and 3CRR samples respectively. 

Using the 11.3\moo\ feature as a clean identifier, we find 8 more objects with low equivalent width (EW) or lower S/N PAH detections (PKS0806-10, PKS1151-34, PKS1814-63, PKS1934-63, 3C33, 3C236, 3C321, 3C326).
The detection strategy used to identify PAH emission in these objects was based first on checking that the candidate feature was detected in both nod positions of the IRS spectra (where both were available).  Second, we modeled each spectrum with PAHFIT, including or excluding the PAH emission components. After subtracting the PAHFIT model that did not include the PAH components from the spectrum, we were able to analyze the resulting residual spectrum for evidence of PAH emission. 

\begin{figure*}
\epsscale{2}
\plottwo{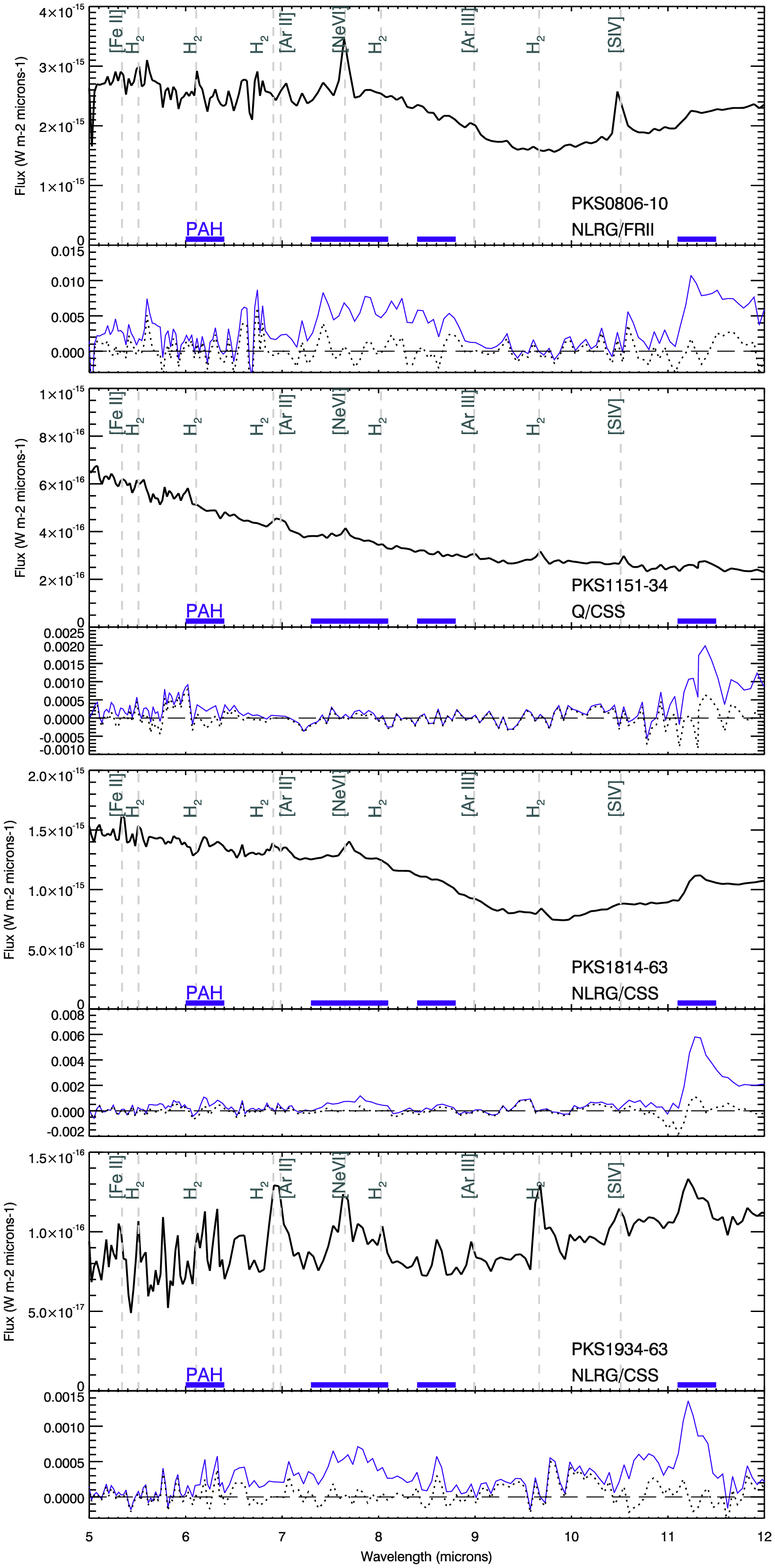}{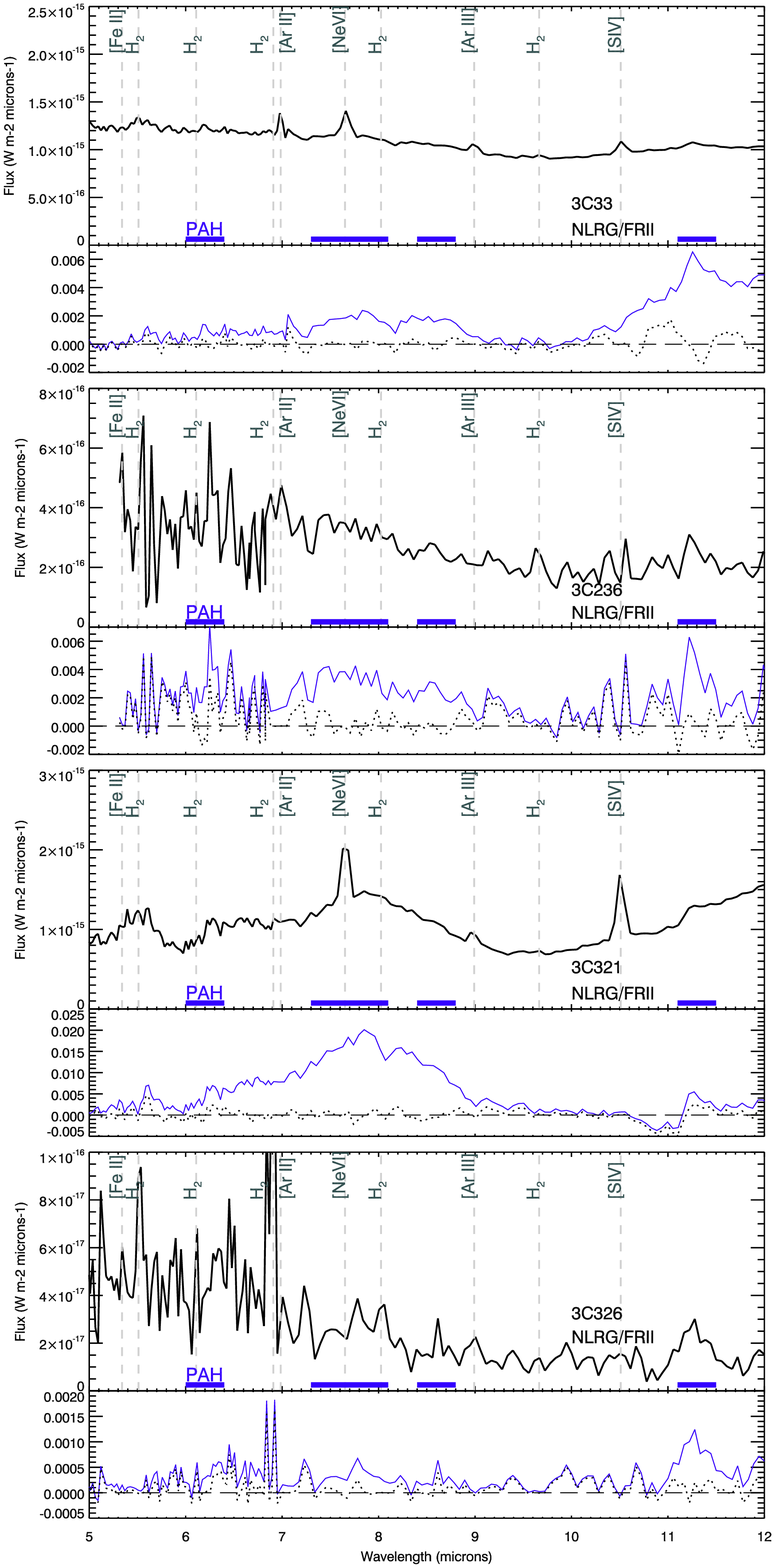}
\caption{: Magnified IRS spectra focussing on the 5-12\moo\ band presenting the objects deemed to have low EW PAH. Below each spectrum are plotted the residuals from the PAHFIT fit , the solid line representing the PAHFIT fit without PAH and the dotted line representing a fit with PAH component. See Section \ref{sec:pah} for more details. \label{fig:PAH_low} }
\end{figure*}

To emphasize these lower level detections we present enlargements of the spectra in Figure \ref{fig:PAH_low}. Below each spectrum in Figure \ref{fig:PAH_low} we plot the residuals from the PAHFIT model, where the solid line represents the residuals from subtracting the spectrum from the PAHFIT model that did not include PAH in the fit, and the dotted line represents residuals of the PAHFIT model that included the PAH features. Four objects (PKS1814-63, PKS1934-61, 3C33, 3C326) have a prominent 11.3\moo\ PAH feature fitted well by the PAHFIT model. In addition, low EW PAH features at 7.7\moo\ appear to be detected. Therefore we argue that there is good evidence in these 4 objects for PAH emission. 

A further four objects (PKS0806-10, PKS1151-34, 3C236, 3C321) have lower EW or irregular PAH detections at 11.3\moo.  Because the 11.3\moo\ feature also lies at the red end of the 10\moo\ silicate absorption feature, potentially the wing of this feature could be mistaken as a low EW signature of PAH. However, for PKS0806-10 and 3C321
the inflection at 11.3\moo\  is much sharper than expected for the wing of the 10\moo\ silicate feature; in both objects the feature remains following subtraction of the PAHFIT model that includes silicate absorption but excludes PAH features. Therefore we regard the detection of PAH features in these two objects as secure. 

PKS1151-34 also appears to have a prominent 11.3\moo\ feature, albeit with an irregular shape. The irregular shape is due to the fact that the feature coincides with the join between the SL and LL parts of the spectrum which hinders a more definitive detection. 

Finally we note that 3C236 also shows a possible 11.3\moo\ feature detected at low S/N in its low resolution IRS spectrum (see Figure 9). Although this detection in not secure based on the low resolution IRS spectrum alone (taken in mapping mode), it has been confirmed with a much higher S/N spectra taken using the high resolution mode of Spitzer/IRS (P. Guillard, private communication).

\subsection{PAH emission strength}

PAHFIT uses a Drude profile method to recover the full strength of the PAH dust emission features and blends. Drude profiles are known to recover the flux in PAH features more accurately than methods which estimate the underlying continuum using line segments or spline curve fits, because they better account for the power in the wings of the broad emission profiles. PAHFIT also corrects the PAH fluxes for the corresponding effects of the silicate absorption model fitted to the data. Converting these fluxes into luminosities gives an opportunity to compare the relative strengths of the emission features for all the objects with IRS spectra. 

Our PAHFIT model fits the detected 11.3\moo\ PAH emission features and returns flux values for all the objects in the 2Jy and 3CRR samples with PAH detections. However, in the case of those objects in which the PAH features are not detected, it is necessary to derive upper limits. To do this in a robust fashion, a scaled IRS spectrum of PKS0915-11 (Hydra A), which is dominated by prominent PAH features, was added to the spectrum of each object without evidence for PAH emission. The scaling factor was varied until a PAH feature at 11.3\moo\ was just detected in a visual inspection of the combined spectrum. Again the 11.3\moo\ feature was used, as this is not contaminated by fine structure lines. Multiplying the measured PAH flux for PKS0915-11 by the scaling factor then gives a robust upper limit on the flux of the PAH. The PAH fluxes and upper limits are presented in Tables \ref{tbl-pah} and \ref{tbl-pah3CRR}.

\begin{figure}[h]
\epsscale{1}
\plotone{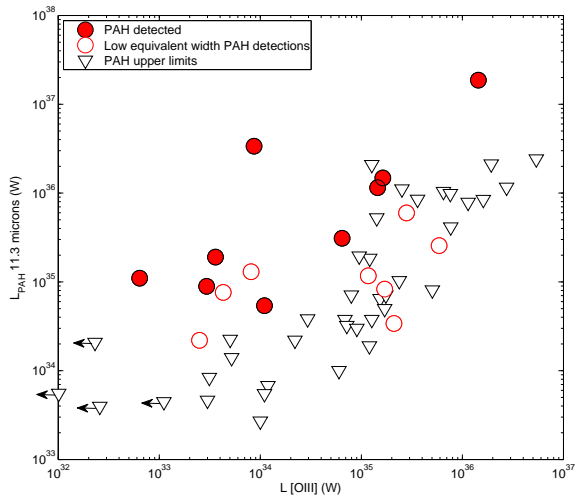}
\caption{: Plot of  $L_{PAH 11.3}$ versus $L_{[\rm{OIII}]}$ \label{fig:pah} }
\end{figure}

Using these upper limits we can test whether the objects with detected PAH emission have a unique signature for starbursts, in that the luminosity of the detections is greater than the upper limits, or whether significant PAH emission could exist in all the spectra but may not be detected due to low S/N data. In Figure \ref{fig:pah} we plot the 11.3\moo\ luminosities against the AGN power indicator $L_{[\rm{OIII}]}$. From this figure it is clear that, on average, the PAH luminosities of objects with high EW PAH detections are higher than the majority of  PAH upper limits for a comparable $L_{[OIII]}$ i.e. AGN power.  

The difference in Figure \ref{fig:pah} between the PAH detections and the upper limits is clearest for low power AGN, but becomes less clear for higher power AGN. Two factors may contribute to this trend. First, because of limits on the exposure times, some of the more powerful, distant objects have lower S/N spectra than their lower power counterparts at low redshifts, leading to higher upper limit flux values. In this context it is notable that the group of upper limits around $L_{PAH}\sim10^{36}$ W, $L_{[OIII]}\sim10^{35}$ W --- which includes PKS0409-75, PKS1306-09, and PKS0252-72 --- includes some of the highest redshift objects in the 2Jy sample ($z > 0.5$); these objects
have relatively low S/N IRS spectra. Second, the relatively stronger AGN dust continuum emission in the more powerful objects could also lead to higher upper limits for the PAH emission.

The objects with low EW PAH detections are also marked in Figure \ref{fig:pah}. In general, they occupy the same region of the plot as the upper limits, but tend to lower PAH luminosities compared to the higher EW PAH detections.

From these data alone it is impossible to entirely rule out RSFA activity, at some level, in the objects with PAH upper limits. However, the position of the upper limits for lower luminosity AGN argues against powerful starburst activity in these objects. 

\begin{figure}[h]
\epsscale{1}
\plotone{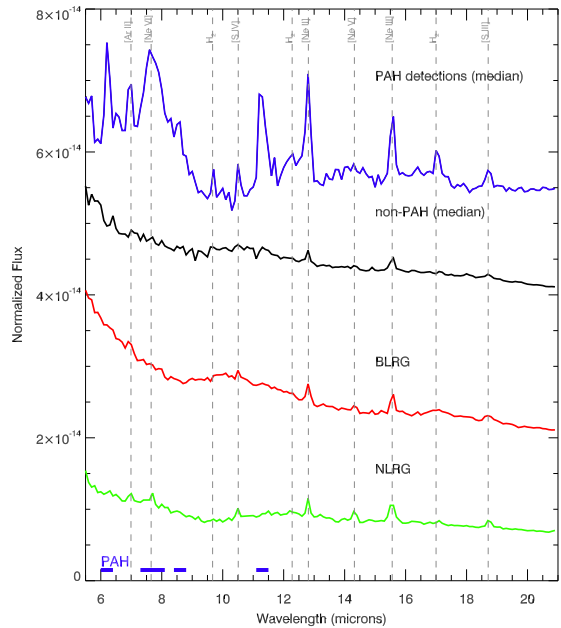}
\caption{: Median spectra for the combined 2Jy and 3CRR samples; objects with PAH detections (blue), objects without PAH detections (black), BLRG without PAH detections (red), NLRG without PAH detections (green). The median spectra were normalized to 1Jy at 20\moo\ and offset in the plot in order to compare them. \label{median_pah} }
\end{figure}

\begin{deluxetable}{l@{\hspace{0mm}}c@{\hspace{3mm}}c@{\hspace{3mm}}c@{\hspace{0mm}}c@{\hspace{0mm}}l@{\hspace{0mm}}r}
\tabletypesize{\scriptsize}
\tablecaption{The 2Jy Sample - PAH Flux Data\label{tbl-pah}}
\tablewidth{0pt}
\tablehead{
\colhead{PKS Name}{\hspace{0mm}} & \colhead{other}{\hspace{0mm}} &\colhead{z}{\hspace{0mm}} &\colhead{Opt.}{\hspace{0mm}} &\colhead{Rad.}{\hspace{0mm}}  & \colhead{$S_{11.3}$(W/$m^2$)}{\hspace{-3mm}}& \colhead{70/24}{\hspace{-3mm}} 
}
\startdata
0023$-$26	&	\phantom{a}		&	\phantom{a}	0.322	&	NLRG	&	CSS	&		$(3.4\pm0.1)\times10^{-17}$	&	\phantom{a}		12.2	$\pm$	2.1	\\
0034$-$01	&	\phantom{a}	3C15	&	\phantom{a}	0.073	&	WLRG	&	FRII	&	$<$	6.6E-18	&	\phantom{a}		2.4	$\pm$	0.3	\\
0035$-$02	&	\phantom{a}	3C17	&	\phantom{a}	0.220	&	BLRG	&	(FRII)	&	$<$	1.3E-17	&	\phantom{a}		1.9	$\pm$	0.4	\\
0038$+$09	&	\phantom{a}	3C18	&	\phantom{a}	0.188	&	BLRG	&	FRII	&	$<$	6.6E-18	&	\phantom{a}		1.2	$\pm$	0.2	\\
0039$-$44	&	\phantom{a}		&	\phantom{a}	0.346	&	NLRG	&	FRII	&	$<$	2.0E-17	&	\phantom{a}		2.1	$\pm$	0.1	\\
0043$-$42	&	\phantom{a}		&	\phantom{a}	0.116	&	WLRG	&	FRII	&	$<$	6.6E-18	&	\phantom{a}		0.9	$\pm$	0.3	\\
0105$-$16	&	\phantom{a}	3C32	&	\phantom{a}	0.400	&	NLRG	&	FRII	&	$<$	2.0E-17	&	\phantom{a}	$<$	1			\\
0117$-$15	&	\phantom{a}	3C38	&	\phantom{a}	0.565	&	NLRG	&	FRII	&	$<$	6.6E-18	&	\phantom{a}		3.3	$\pm$	0.3	\\
0213$-$13	&	\phantom{a}	3C62	&	\phantom{a}	0.147	&	NLRG	&	FRII	&	$<$	6.6E-18	&	\phantom{a}		0.9	$\pm$	0.1	\\
0235$-$19	&	\phantom{a}	OD-159	&	\phantom{a}	0.620	&	BLRG	&	FRII	&	$<$	1.3E-17	&	\phantom{a}		1.3	$\pm$	0.2	\\
0252$-$71	&	\phantom{a}		&	\phantom{a}	0.566	&	NLRG	&	CSS	&	$<$	6.6E-18	&	\phantom{a}	$<$	3			\\
0347$+$05	&	\phantom{a}		&	\phantom{a}	0.339	&	WLRG	&	FRII	&		$(8.9\pm0.1)\times10^{-17}$	&	\phantom{a}		8.8	$\pm$	1.3	\\
0349$-$27	&	\phantom{a}		&	\phantom{a}	0.066	&	NLRG	&	FRII	&	$<$	6.6E-18	&	\phantom{a}		4.8	$\pm$	0.2	\\
0404$+$03	&	\phantom{a}	3C105	&	\phantom{a}	0.089	&	NLRG	&	FRII	&	$<$	2.0E-17	&	\phantom{a}		2.3	$\pm$	0.1	\\
0409$-$75	&	\phantom{a}		&	\phantom{a}	0.693	&	NLRG	&	FRII	&	$<$	9.9E-18	&	\phantom{a}		7.3	$\pm$	1.9	\\
0442$-$28	&	\phantom{a}		&	\phantom{a}	0.147	&	NLRG	&	FRII	&	$<$	6.6E-18	&	\phantom{a}		1.4	$\pm$	0.2	\\
0620$-$52	&	\phantom{a}		&	\phantom{a}	0.051	&	WLRG	&	FRI	&	$<$	6.6E-18	&	\phantom{a}		10.5	$\pm$	0.4	\\
0625$-$35	&	\phantom{a}	OH-342	&	\phantom{a}	0.055	&	WLRG/BLLac	&	FRI	&	$<$	6.6E-18	&	\phantom{a}		1.8	$\pm$	0.1	\\
0625$-$53	&	\phantom{a}		&	\phantom{a}	0.054	&	WLRG	&	FRII	&	$<$	6.6E-18	&	\phantom{a}	$<$	6			\\
0806$-$10	&	\phantom{a}	3C195	&	\phantom{a}	0.110	&	NLRG	&	FRII	&		$(1.8\pm1.5)\times10^{-17}$	&	\phantom{a}		1.9	$\pm$	0.01	\\
0859$-$25	&	\phantom{a}		&	\phantom{a}	0.305	&	NLRG	&	FRII	&	$<$	6.6E-18	&	\phantom{a}		0.9	$\pm$	0.3	\\
0915$-$11	&	\phantom{a}	Hydra A	&	\phantom{a}	0.054	&	WLRG	&	FRI	&		$(1.1\pm0.1)\times10^{-16}$	&	\phantom{a}		12.9	$\pm$	0.6	\\
0945$+$07	&	\phantom{a}	3C227	&	\phantom{a}	0.086	&	BLRG	&	FRII	&	$<$	4.0E-17	&	\phantom{a}		0.4	$\pm$	0.1	\\
1136$-$13	&	\phantom{a}		&	\phantom{a}	0.554	&	Q	&	FRII	&	$<$	2.0E-17	&	\phantom{a}		1.7	$\pm$	0.2	\\
1151$-$34	&	\phantom{a}		&	\phantom{a}	0.258	&	Q	&	CSS	&		$(3.0\pm0.1)\times10^{-17}$	&	\phantom{a}		3.2	$\pm$	0.2	\\
1306$-$09	&	\phantom{a}		&	\phantom{a}	0.464	&	NLRG	&	CSS	&	$<$	6.6E-18	&	\phantom{a}		4.7	$\pm$	0.5	\\
1355$-$41	&	\phantom{a}		&	\phantom{a}	0.313	&	Q	&	FRII	&	$<$	1.3E-17	&	\phantom{a}		1.2	$\pm$	0.04	\\
1547$-$79	&	\phantom{a}		&	\phantom{a}	0.483	&	BLRG	&	FRII	&	$<$	1.3E-17	&	\phantom{a}		2.4	$\pm$	0.2	\\
1559$+$02	&	\phantom{a}	3C327	&	\phantom{a}	0.104	&	NLRG	&	FRII	&	$<$	2.6E-17	&	\phantom{a}		1.9	$\pm$	0.02	\\
1602$+$01	&	\phantom{a}	3C327.1	&	\phantom{a}	0.462	&	BLRG	&	FRII	&	$<$	1.3E-17	&	\phantom{a}		1.6	$\pm$	0.3	\\
1648$+$05	&	\phantom{a}	Herc A	&	\phantom{a}	0.154	&	WLRG	&	FRI	&		--	&	\phantom{a}	$<$	9			\\
1733$-$56	&	\phantom{a}		&	\phantom{a}	0.098	&	BLRG	&	FRII	&		$(1.3\pm0.1)\times10^{-16}$	&	\phantom{a}		5.2	$\pm$	0.1	\\
1814$-$63	&	\phantom{a}		&	\phantom{a}	0.063	&	NLRG	&	CSS	&		$(8.2\pm0.5)\times10^{-17}$	&	\phantom{a}		2.3	$\pm$	0.0	\\
1839$-$48	&	\phantom{a}		&	\phantom{a}	0.112	&	WLRG	&	FRI	&	$<$	6.6E-18	&	\phantom{a}		3.5	$\pm$	0.9	\\
1932$-$46	&	\phantom{a}		&	\phantom{a}	0.231	&	BLRG	&	FRII	&	$<$	6.6E-18	&	\phantom{a}		7.1	$\pm$	0.7	\\
1934$-$63	&	\phantom{a}		&	\phantom{a}	0.183	&	NLRG	&	GPS	&		$(1.3\pm0.1)\times10^{-17}$	&	\phantom{a}		1.1	$\pm$	0.1	\\
1938$-$15	&	\phantom{a}		&	\phantom{a}	0.452	&	BLRG	&	FRII	&	$<$	1.3E-17	&	\phantom{a}		2.9	$\pm$	0.6	\\
1949$+$02	&	\phantom{a}	3C403	&	\phantom{a}	0.059	&	NLRG	&	FRII	&	$<$	4.0E-17	&	\phantom{a}		1.8	$\pm$	0.02	\\
1954$-$55	&	\phantom{a}		&	\phantom{a}	0.060	&	WLRG	&	FRI	&	$<$	6.6E-18	&	\phantom{a}		3.3	$\pm$	1.1	\\
2135$-$14	&	\phantom{a}		&	\phantom{a}	0.200	&	Q	&	FRII	&		--	&	\phantom{a}		1.1	$\pm$	0.05	\\
2135$-$20	&	\phantom{a}	OX-258	&	\phantom{a}	0.635	&	BLRG	&	CSS	&		$(1.1\pm0.1)\times10^{-16}$	&	\phantom{a}		8.7	$\pm$	1.0	\\
2211$-$17	&	\phantom{a}	3C444	&	\phantom{a}	0.153	&	WLRG	&	FRII	&		--	&	\phantom{a}	$<$	18			\\
2221$-$02	&	\phantom{a}	3C445	&	\phantom{a}	0.057	&	BLRG	&	FRII	&	$<$	6.6E-17	&	\phantom{a}		0.8	$\pm$	0.02	\\
2250$-$41	&	\phantom{a}		&	\phantom{a}	0.310	&	NLRG	&	FRII	&	$<$	2.6E-18	&	\phantom{a}		1.9	$\pm$	0.2	\\
2314$+$03	&	\phantom{a}	3C459	&	\phantom{a}	0.220	&	NLRG	&	FRII	&		$(1.1\pm0.1)\times10^{-16}$	&	\phantom{a}		10.3	$\pm$	0.1	\\
2356$-$61	&	\phantom{a}		&	\phantom{a}	0.096	&	NLRG	&	FRII	&	$<$	1.3E-17	&	\phantom{a}		1.8	$\pm$	0.1	\\
 \enddata

\tablecomments{ Table presenting luminosities derived from the fluxes extracted using the
PAHFIT program for all 2Jy objects with useful {\it IRS}
data. Uncertainties are derived from the PAHFIT model fit. See Table\ref{tbl-ratio} for ratios of 11.3/6.2\moo\ and 11.3/7.7\moo\ fluxes. PKS0347+05 had been classified in our previous work as a BLRG (T07, D08, D09), however, we have since determined that the broad-line component comes from a companion galaxy and not the radio galaxy, therefore it is now classified as a WLRG.  }
\end{deluxetable}

\begin{deluxetable}{l@{\hspace{0mm}}c@{\hspace{3mm}}c@{\hspace{0mm}}c@{\hspace{0mm}}l@{\hspace{0mm}}r}
\tabletypesize{\scriptsize}
\tablecaption{The 3CRR Sample -  PAH Flux Data\label{tbl-pah3CRR}}
\tablewidth{0pt}
\tablehead{
\colhead{Name}{\hspace{0mm}} & \colhead{z}{\hspace{0mm}} &\colhead{Opt.}{\hspace{0mm}} &\colhead{Rad.}{\hspace{0mm}}  & \colhead{$S_{11.3}$(W/$m^2$)}{\hspace{-3mm}}& \colhead{70/24}{\hspace{-3mm}} 
}
\startdata
3C33	&	\phantom{a}	0.060	&	NLRG	&	FRII	&\phantom{a}		$(8.9\pm0.6)\times10^{-17}$	&		1.5	$\pm$	0.03	\\
3C35	&	\phantom{a}	0.067	&	WLRG	&	FRII	&\phantom{a}		--	&\phantom{a}		21.0	$\pm$	8.6	\\
3C98	&	\phantom{a}	0.030	&	NLRG	&	FRII	&\phantom{a}	$<$	1.3E-17	&		0.8	$\pm$	0.1	\\
DA240	&	\phantom{a}	0.036	&	WLRG	&	FRII	&\phantom{a}		--	&		8.3	$\pm$	1.5	\\
3C192	&	\phantom{a}	0.060	&	NLRG	&	FRII	&\phantom{a}	$<$	2.6E-17	&		2.4	$\pm$	1.1	\\
4C73.08	&	\phantom{a}	0.058	&	NLRG	&	FRII	&\phantom{a}		--	&		0.5	$\pm$	0.1	\\
3C236	&	\phantom{a}	0.101	&	WLRG	&	FRII	&\phantom{a}		$(5.4\pm0.1)\times10^{-17}$	&		3.7	$\pm$	0.3	\\
3C277.3	&	\phantom{a}	0.085	&	WLRG	&	FRI/FRII	&\phantom{a}		--	&		2.1	$\pm$	0.4	\\
3C285	&	\phantom{a}	0.079	&	NLRG	&	FRII	&\phantom{a}		$(1.3\pm0.1)\times10^{-16}$	&		4.3	$\pm$	0.1	\\
3C293	&	\phantom{a}	0.045	&	WLRG	&	FRI/FRII	&\phantom{a}		$(2.3\pm0.1)\times10^{-16}$	&		9.7	$\pm$	0.2	\\
3C305	&	\phantom{a}	0.042	&	NLRG	&	FRII/CSS	&\phantom{a}		$(1.4\pm0.3)\times10^{-16}$	&		7.1	$\pm$	0.1	\\
3C321	&	\phantom{a}	0.096	&	NLRG	&	FRII	&\phantom{a}		$(1.5\pm0.1)\times10^{-17}$	&		3.4	$\pm$	0.02	\\
3C326	&	\phantom{a}	0.090	&	NLRG	&	FRII	&\phantom{a}		$(1.1\pm0.1)\times10^{-17}$	&\phantom{a}	$<$	13			\\
3C382	&	\phantom{a}	0.058	&	BLRG	&	FRII	&\phantom{a}	$<$	1.3E-17	&		0.6	$\pm$	0.04	\\
3C388	&	\phantom{a}	0.092	&	WLRG	&	FRII	&\phantom{a}	$<$	6.6E-18	&	$<$	4			\\
3C390.3	&	\phantom{a}	0.056	&	BLRG	&	FRII	&\phantom{a}	$<$	2.6E-17	&		0.8	$\pm$	0.01	\\
3C452	&	\phantom{a}	0.081	&	NLRG	&	FRII	&\phantom{a}	$<$	6.6E-18	&		1.0	$\pm$	0.1	\\	\enddata

\tablecomments{Table presenting luminosities derived from the fluxes extracted using the
PAHFIT program for all 3CRR the objects with useful {\it IRS}
data.  Note that 3C403 and 3C445 overlap between the two samples; see Table \ref{tbl-pah} for PAH data for these latter objects.}
\end{deluxetable}

By combining the spectra that have no evidence for PAH features, it is possible to create a spectrum with high signal-to-noise to investigate the possibility of low level PAH emission in the objects without individual detections. We present median spectra for the objects without PAH detections in the combined 2Jy and 3CRR sample in Figure \ref{median_pah}. These median spectra were created by re-sampling the spectral data in wavelength bins and normalizing all the spectra to have the same flux  at 20\moo. While the median spectrum of the PAH identified objects shows the characteristic PAH emission dominating the spectrum, the median spectrum for objects in which we have not individually identified PAH shows no evidence for PAH features. 

Also in Figure \ref{median_pah} we present median spectra for objects without individually identified PAH features divided into BLRG and NLRG. We can use these median spectra to test the idea that a stronger AGN continuum in BLRG may mask low level PAH emission, which in turn may be more easily detected in NLRG. The test is again negative, revealing no evidence for PAH emission in the NLRG. 

\subsection{Comparison with optical star formation indicators}
\label{sec:star}

Careful spectral synthesis modeling of deep optical spectra for the 2Jy (\citealp{tadhunter02}; \citealp{wills04, wills08}; \citealp{holt07}) and 3CRR (see \citet{dicken10} and references therein) samples has allowed identification of objects with evidence for RSFA at optical wavelengths. 

Considering the objects with high EW PAH detections that also have deep optical spectra available,  it is notable that \emph{all} 7 of these objects show evidence of young stellar populations in their optical spectra. In addition, we have identified RSFA through PAH detection in 2 objects (PKS0347+05, PKS1733-56) for which the contamination by direct AGN emission components precluded the identification of any young stellar populations at optical wavelengths. 

Out of the 12 objects with good optical evidence for RSFA and IRS spectra, only 3 objects show no evidence for PAH features (PKS0409-75, PKS0620-52, PKS1932-46). The relatively low S/N of the spectra in the region of the 11.3\moo\ PAH feature for these 3 objects (S/N = 2, 13 and 4 respectively, compared with the sample average of 24) could help to explain why no PAH emission is detected. 

Overall the optical and PAH evidence for RSFA correlates well. The correlation is unlikely to be perfect since the two star formation indicators do not necessarily sample star formation activity on the same spatial scales, or at the same evolutionary stages of the starbursts. For example, in a significant subset of the radio galaxies with optically identified young stellar populations we are detecting post-starburst stellar populations that are observed a significant period ($\sim0.1 - 2$~Gyr) after   the peak of merger induced star formation activity (\citealp{tadhunter05,tadhunter11}).  In addition,  identification of RSFA at optical wavelengths is not possible in some objects due to a strong AGN continuum that masks these signatures. Equally, PAH identification may be hindered by low S/N spectra and/or calibration errors such as noted for 3C15 (see Section \ref{sec:2}). 

Combining the objects that have high EW PAH detections (9) and the objects that have clear optically identified RSFA (12) makes a total of 14 objects  (22\%) with strong evidence for RSFA in the combined 2Jy and 3CRR sample. 

Next we consider the 6 objects with low EW PAH detections that also have deep optical spectra available. Only 2 of these objects -- 3C236, 3C321 -- have evidence for RSFA at optical wavelengths. The low EW PAH detections in the remaining 4 objects -- PKS0806-10, PKS1934-63, 3C33 and 3C326 -- may represent a low level of star formation that could not be detected at optical wavelengths. However, it is also important to consider that the low EW PAH emission may originate from ISM diffuse PAH emission rather than be associated with RSFA \citep{kaneda08}, particularly for objects with low AGN continuum strength such as 3C326. 

Finally, we return to the question posed in the introduction of whether there is a substantial population of radio galaxies in which the RSFA is missed at optical wavelengths due to dust obscuration. We find that, of the 36 objects that have good optical spectroscopic information and show no sign of RSFA in their optical spectra, only the 4  objects (11\%) discussed above -- PKS0806-10, PKS1934-63, 3C33 and 3C326 -- have PAH features detected in their mid-IR spectra; and these are all low EW detections.  Therefore, the incidence of objects in our samples that have obscured RSFA that is not already detected at optical wavelengths is relatively minor.

\subsection{Mid- to far-IR color}
\label{sec:color}

\begin{figure*}
\epsscale{2}
\plotone{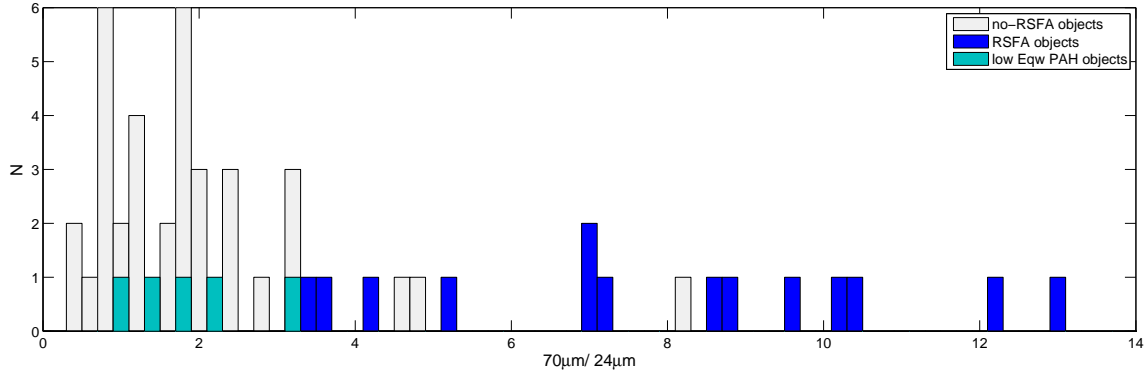}
\caption{: Histogram of  70\moo/24\moo\  color for the  combined 2Jy and 3CRR sample. Note that the figure does not include the 5 objects in the 2Jy sample and the 2 objects in the 3CRR sample with upper limits on their 70\moo\ flux.  \label{fig:hist} }
\end{figure*}

\noindent
A common method for identifying star formation activity involves using the mid- to far-IR (MFIR) color. Both the 2Jy and 3CRR samples have complete Spitzer MIPS data which can be used to investigate the F(70)/F(24) flux ratio (D09, D10). Empirically, starburst galaxies are known to be associated with cool MFIR colors, as expected in the case of illumination of extended dust structures by spatially distributed star forming regions. Unlike PAH emission, the MFIR color diagnostic is not a clean identifier of RSFA, because the far-IR emitting dust can be heated by AGN and/or young stellar populations. However, in our previous work (D09, D10) we have shown that optically identified RSFA objects do tend to have cooler MFIR colors. With the addition of the more complete PAH identified RSFA objects from the 2Jy and 3CRR samples, we can thoroughly test the MFIR color star formation diagnostic for radio-loud AGN. 

Figure \ref{fig:hist} presents a histogram of the MFIR colors for the 2Jy and 3CRR samples, with the RSFA objects indicated. It is immediately clear that all but one of the 11 objects with F(70)/F(24) $>$5 show evidence for RSFA on the basis of optical and/or PAH data, whereas the majority of objects lacking such evidence cluster around MFIR colors of F(70)/F(24) $\approx$ 0.5-2.5. However, there is a mix of objects in the range 2.5$<$F(70)/F(24)$<$5, which may represent a MFIR color transition region between AGN- and RSFA-dominated objects. Objects in this transition region with F(70)/F(24) $\approx$4.5 include PKS0349-27 and PKS1306-09. We note that these latter two objects do not have the high quality spectra required to detect the signatures of RSFA at optical wavelengths. It is also noteworthy that PKS1306-09 is a CSS object that shows a strong far-IR excess, in the sense that it falls well above the $L_{70\mu m}$  vs $L_{[\rm{OIII}]}$ correlation  (see Section \ref{sec:css}). 

{\it All} the objects with either high EW PAH and/or optical evidence for RSFA (highlighted in dark blue in Figure \ref{fig:hist}) show relatively cool MFIR colors with 70\moo/24\moo\ $>$ 3. Therefore this study finds that MFIR color is an excellent indicator of starburst activity in radio galaxies. This is consistent with the results found by \citet{brandl06} for AGN using the F(15)/F(30) ratio, and by \citet{veilleux09a} using the F(60)/F(25) ratio for at large sample of
ULIRGs with varying AGN contribution.  

Of the 8 objects with low EW PAH detections,  5 have been identified in Figure \ref{fig:hist}, while a further two objects (3C236 and 3C321) are already highlighted in this figure as optical RSFA objects (dark blue) with F(70)/F(24) = 3.7 and 3.4 respectively. The eighth low EW PAH object (3C326) is not included in this figure as it has an upper limit on F(70)/F(24). The figure shows that the MFIR colors of the low EW PAH objects are similar to those of objects lacking PAH detections. This result is not surprising assuming that the low EW PAH detections represent a low level of star formation activity or indeed if the PAH has a diffuse ISM emission origin rather than a stellar origin as mentioned in above in Section \ref{sec:star}. It is notable that the two low EW PAH objects with the coolest colors, 3C236 and 3C321, mentioned above, also have optical RSFA detections.

It is also important to consider the possibility of RSFA in the 3 objects from the 2Jy sample and the 4 objects from the 3CRR sample that were not observed/detected with Spitzer/IRS. None of these objects have optical evidence for RSFA. However, two of the closest objects (z$\approx$0.03) 3C35 and DA240 have cool MFIR colors (=21.0$\pm$8.6  and 8.3$\pm$1.5 respectively).  We have excluded 3C35 from this analysis because its measured F(70)/F(24) has a large uncertainty. DA240 is the only object with cool MFIR colors F(70)/F(24)$>$5 without clear optical or PAH identification of RSFA.  On the other hand, this object does show evidence for RSFA in the form of a far-IR excess (see Section \ref{sec:origin1} and Table \ref{tbl-SBdetect}). 

\subsection{Far-IR excess}
\label{sec:origin1}

\begin{figure*}
\epsscale{2}
\plotone{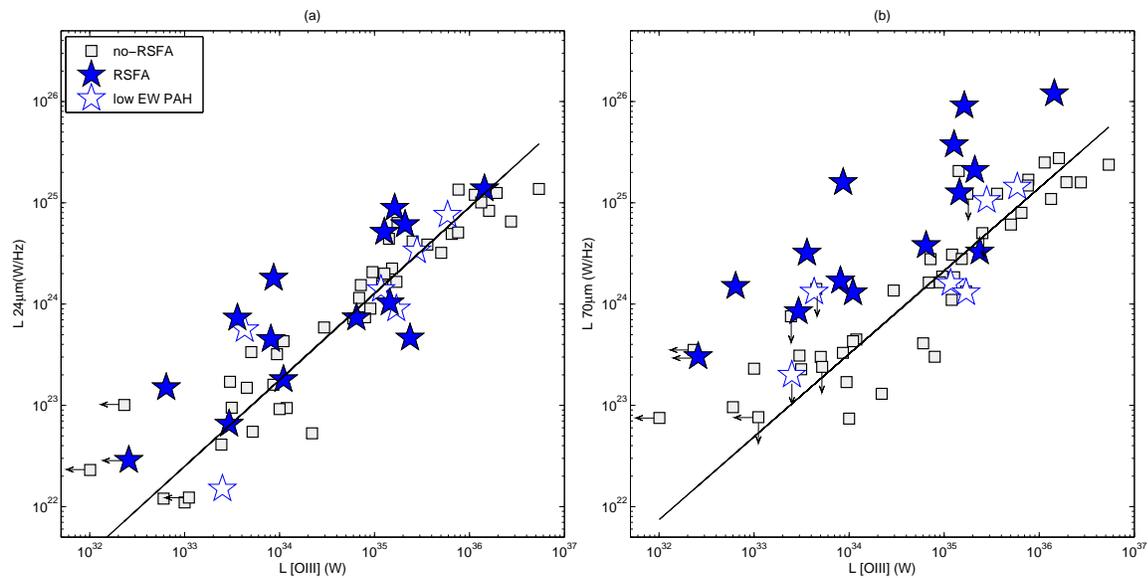}
\caption{: Plot showing $L_{24\mu m}$ and $L_{70\mu m}$  versus $L_{[\rm{OIII}]}$ where those objects identified to contain evidence for RSFA with optical and/or PAH diagnostics are indicated with blue stars. The lines shown are the bisectors for no-RSFA objects of linear least squares fits of $x$ on $y$ and $y$ on $x$. The fit also does not include the 7 objects with upper limits in either [OIII] or 70\moo. The 24\moo, 70\moo\ and [OIII] fluxes and luminosities for the 2Jy sample, can be found in D08 and D09; fluxes for the 3CRR sample are presented in \citet{dicken10}.  \label{fig_corr} }
\end{figure*}

\noindent
Our previous work has shown that the objects with optically identified RSFA have enhanced emission at far-IR wavelengths compared to the majority of the sample objects (T07, D09, D10). Several objects in these previous studies also showed apparent enhancements in their far-IR emission, but optical identification of RSFA was not possible. We are now able to revisit this investigation with the benefit of the IRS data and PAH detections to see if our previous result holds with the new, more complete, RSFA identifications. 

In Figure \ref{fig_corr} we plot  $L_{24\mu m}$ and $L_{70\mu m}$  against $L_{[\rm{OIII}]}$ for the 2Jy and 3CRR samples. As discussed in Section \ref{sec:intro}, $L_{[\rm{OIII}]}$ is a good indicator of the intrinsic power of the AGN. Consequently, we interpret the similar slopes of the correlations\footnote{Spearman partial rank correlation tests presented in D10 show the correlations are intrinsic and do not arise because $L_{[\rm{OIII}]}$ and $L_{MFIR}$ are independently correlated with redshift. The null hypothesis that the variables are unrelated is still rejected at the $>$99.5\% level of significance.} between $L_{24\mu m}$,  $L_{70\mu m}$ and $L_{[\rm{OIII}]}$ seen in Figure \ref{fig_corr} as direct evidence for AGN illumination as the dominant heating mechanism of the thermal mid- to far-IR emitting dust (D09, D10). Considering the spectral diversity seen in the IRS data presented in this paper, the correlation between the mid-IR ($L_{24\mu m}$)  and $L_{[\rm{OIII}]}$ is striking. 

Following the method of our previous investigations we have marked with blue stars in Figure \ref{fig_corr} those objects with high EW PAH detections and/or optical evidence for RSFA in the 2Jy and 3CRR samples. Consistent with our previous studies, there is little difference between the distributions of RSFA and non-RSFA objects in the  $L_{24\mu m}$  vs $L_{[\rm{OIII}]}$ plot (Figure \ref{fig_corr} (a)). In contrast, Figure \ref{fig_corr} (b) clearly shows that all but one (PKS1932-46)\footnote{This object is discussed in detail in D10. Amongst the objects with optically identified RSFA, this object has the weakest evidence for RSFA in its near-nuclear regions, with the evidence confined to the detection of HII region-like emission line ratios in the extended halo of the galaxy \citep{villar_martin05}.} of the 14 objects with optical and/or PAH evidence for RSFA in the 2Jy and 3CRR samples lie above the $L_{70\mu m}$ versus $L_{[\rm{OIII}]}$ correlation defined by the regression line fitted to the no-RSFA objects.

This boosting of the  objects with evidence for RSFA above the correlation between $L_{70\mu m}$ and $L_{[\rm{OIII}]}$  in Figure \ref{fig_corr} --- by more than an order of magnitude in some cases --- provides strong evidence for enhancement of the thermal far-IR emission by RSFA. Conducting a one dimensional Kolmogorov--Smirnov two sample test on the combined sample, comparing the vertical displacements from our fitted regression line in the $L_{70\mu m}$ vs $L_{[\rm{OIII}]}$ plot, we find that we can reject the null hypothesis that the objects with and without independent optical/PAH evidence for RSFA are drawn from the same parent population at a better than 0.01\% level of significance.
Note that, in conducting this analysis, we included upper limits on $L_{70\mu m}$ and $L_{[\rm{OIII}]}$ as actual measurements.

To take into account the upper limits in a more rigorous manner, we also investigated the differences in the distributions of $L_{70\mu m}$ and $L_{[\rm{OIII}]}$ including the upper limits, using survival statistics  in the ASURV package (\citealp{isobe86}; \citealp{lavalley92}) as implemented in IRAF. Specifically, the "twosampt" task, which computes several nonparametric two-sample tests for comparing two or more censored data sets, was used. These tests revealed that the 70\moo\ flux was significantly different in the starburst versus the non-starburst populations. For instance, the Gehan's Generalized Wilcoxon test estimated that the probability ($p$) that the two sets of censored data belonged to the same population was = 0.023 (test statistic = 2.268) while the Logrank test estimates $p$ = 0.013 (test statistic = 2.483). On the other hand, the [OIII] flux was not significantly different between the two populations: the Gehan's Generalized Wilcoxon test gave a test statistic = 0.531 and $p$ = 0.595, while the Logrank test gave a test statistic = 0.123 and $p$ = 0.902.

We have also plotted the objects with low EW PAH detections on Figure \ref{fig_corr}. Many of the low EW PAH objects fall above the $L_{70\mu m}$ versus $L_{[\rm{OIII}]}$ correlation, but the result is not as clear-cut as it is for the high EW PAH objects. It is noteworthy that the low EW PAH object furthest above the fitted regression line in the $L_{70\mu m}$ vs $L_{[\rm{OIII}]}$ correlation -- 3C321 -- has the coolest MFIR colors out of all the low EW PAH objects (F(70)/F(24) = 3.4) and also optical identification of RSFA. This object is indicated with a filled star in Figure \ref{fig_corr}. 

In terms of whether the far-IR excess alone is a good indicator of RSFA, we find that all 13  of the objects that fall more than 0.5 dex (factor of 3) above the $L_{70\mu m}$ and $L_{[\rm{OIII}]}$ 
regression line also show evidence for significant RSFA based on at least two of the other indicators. We conclude that far-IR excess is as useful an indicator of RSFA in radio galaxies as MFIR color.

\section{Discussion}
\label{sec:disc}

\subsection{Star formation activity in radio-loud AGN}

Our multi-wavelength dataset allows us to estimate the maximum percentage of objects that show evidence for RSFA in the combined 2Jy and 3CRR sample by considering all the main optical and infrared diagnostics of star formation activity. In order to summarize the results we have collated the data for the various RSFA identifiers in Table \ref{tbl-SBdetect}. Several objects are not complete in terms of the available optical, PAH, far-IR excess and color tests of RSFA (indicated by a ``U'' in the Table). However, {\t all} the objects have at least two  separate pieces of RSFA diagnostic information. We further note that, of the 17 objects with optical, low or high EW PAH, color-based or infrared excess identification of RSFA, 7 objects are identified as RSFA based on all four diagnostics, and a further 9 objects on the basis of two diagnostics.

Considering now the full  2Jy+3CRR sample of 63 objects: 19\%  (12 objects) show unambiguous spectroscopic evidence for RSFA at optical wavelengths; 21\% (13) have cool MFIR colors ($F(70/F(24) > 5$); 14\% (9) have high EW PAH emission; 22\% (14) lie more than 0.5 dex (a factor of 3) above the regression line fitted to the $L_{70\mu m}$ versus $L_{[\rm{OIII}]}$ correlation. Overall we find that 27\% (17) objects show at least one of these indicators. Including in the analysis the objects with low EW PAH detections (5 additional objects not identified using the other diagnostics), we estimate that the maximum percentage of radio-loud AGN showing evidence for RSFA in the combined 2Jy and 3CRR sample is 35\%.

This combined result from the various diagnostic methods provides conclusive evidence that a one to one connection between RSFA and AGN activity does \emph{not} exist for the majority of intermediate redshift radio-loud AGN. This conclusion is consistent with our previous investigations, as well as other published Spitzer studies in the literature (\citealp{tadhunter07}; D08; D09; \citealp{cleary07}; \citealp{shi07}; \citealp{fu09}) 

\begin{deluxetable}{l@{\hspace{0mm}}l@{\hspace{0mm}}c@{\hspace{0mm}}c@{\hspace{0mm}}c@{\hspace{0mm}}c@{\hspace{0mm}}l@{\hspace{0mm}}l@{\hspace{0mm}}c@{\hspace{0mm}}c@{\hspace{0mm}}c@{\hspace{0mm}}c}
\tabletypesize{\scriptsize}
\tablecaption{Recent star formation activity detections in the 2Jy and 3CRR samples \label{tbl-SBdetect}}
\tablewidth{0pt}
\tablehead{
\colhead{Name}{\hspace{0mm}} & \colhead{other}{\hspace{0mm}} & \colhead{Optical}{\hspace{-3mm}} &\colhead{PAH}{\hspace{-3mm}} &\colhead{Color} &\colhead{Far-IR}  &\colhead{Name}{\hspace{0mm}} & \colhead{other}{\hspace{0mm}} & \colhead{Optical}{\hspace{-3mm}} &\colhead{PAH}{\hspace{-3mm}} &\colhead{Color} &\colhead{Far-IR} 
}
\startdata
0023$-$26	&	\phantom{a}		&	\tick	&	\tick	&	\tick	&	\tick	&	1814$-$63	&	\phantom{a}		&	U	&	low	&	--	&	\tick	\\
0034$-$01	&	\phantom{a}	3C15	&	--	&	--	&	--	&	--	&	1839$-$48	&	\phantom{a}		&	--	&	--	&	--	&	--	\\
0035$-$02	&	\phantom{a}	3C17	&	--	&	--	&	--	&	--	&	1932$-$46	&	\phantom{a}		&	\tick	&	--	&	\tick	&	--	\\
0038$+$09	&	\phantom{a}	3C18	&	--	&	--	&	--	&	--	&	1934$-$63	&	\phantom{a}		&	--	&	low	&	--	&	--	\\
0039$-$44	&	\phantom{a}		&	--	&	--	&	--	&	--	&	1938$-$15	&	\phantom{a}		&	--	&	--	&	--	&	--	\\
0043$-$42	&	\phantom{a}		&	--	&	--	&	--	&	--	&	1949$+$02	&	\phantom{a}	3C403	&	--	&	--	&	--	&	--	\\
0105$-$16	&	\phantom{a}	3C32	&	--	&	--	&	U	&	--	&	1954$-$55	&	\phantom{a}		&	--	&	--	&	--	&	--	\\
0117$-$15	&	\phantom{a}	3C38	&	--	&	--	&	--	&	--	&	2135$-$14	&	\phantom{a}		&	U	&	U	&	--	&	--	\\
0213$-$13	&	\phantom{a}	3C62	&	--	&	--	&	--	&	--	&	2135$-$20	&	\phantom{a}	OX--258	&	\tick	&	\tick	&	\tick	&	\tick	\\
0235$-$19	&	\phantom{a}	OD--159	&	--	&	--	&	--	&	--	&	2211$-$17	&	\phantom{a}	3C444	&	--	&	U	&	U	&	--	\\
0252$-$71	&	\phantom{a}		&	--	&	--	&	U	&	--	&	2221$-$02	&	\phantom{a}	3C445	&	--	&	--	&	--	&	--	\\
0347$+$05	&	\phantom{a}		&	U	&	\tick	&	\tick	&	\tick	&	2250$-$41	&	\phantom{a}		&	--	&	--	&	--	&	--	\\
0349$-$27	&	\phantom{a}		&	U	&	--	&	--	&	--	&	2314$+$03	&	\phantom{a}	3C459	&	\tick	&	\tick	&	\tick	&	\tick	\\
0404$+$03	&	\phantom{a}	3C105	&	--	&	--	&	--	&	--	&	2356$-$61	&	\phantom{a}		&	--	&	--	&	--	&	--	\\
0409$-$75	&	\phantom{a}		&	\tick	&	--	&	\tick	&	\tick	&	3C33	&	\phantom{a}		&	--	&	low	&	--	&	--	\\
0442$-$28	&	\phantom{a}		&	--	&	--	&	--	&	--	&	3C35	&	\phantom{a}		&	--	&	U	&	U	&	--	\\
0620$-$52	&	\phantom{a}		&	\tick	&	--	&	\tick	&	--	&	3C98	&	\phantom{a}		&	--	&	--	&	--	&	--	\\
0625$-$35	&	\phantom{a}	OH--342	&	--	&	--	&	--	&	--	&	DA240	&	\phantom{a}		&	--	&	U	&	\tick	&	\tick	\\
0625$-$53	&	\phantom{a}		&	--	&	--	&	U	&	--	&	3C192	&	\phantom{a}		&	--	&	--	&	--	&	--	\\
0806$-$10	&	\phantom{a}	3C195	&	--	&	low	&	--	&	--	&	4C73.08	&	\phantom{a}		&	--	&	U	&	--	&	--	\\
0859$-$25	&	\phantom{a}		&	--	&	--	&	--	&	--	&	3C236	&	\phantom{a}		&	\tick	&	low	&	--	&	\tick	\\
0915$-$11	&	\phantom{a}	Hydra A	&	\tick	&	\tick	&	\tick	&	\tick	&	3C277.3	&	\phantom{a}		&	--	&	U	&	--	&	--	\\
0945$+$07	&	\phantom{a}	3C227	&	U	&	--	&	--	&	--	&	3C285	&	\phantom{a}		&	\tick	&	\tick	&	\tick	&	\tick	\\
1136$-$13	&	\phantom{a}		&	U	&	--	&	--	&	--	&	3C293	&	\phantom{a}		&	\tick	&	\tick	&	\tick	&	\tick	\\
1151$-$34	&	\phantom{a}		&	U	&	low	&	--	&	--	&	3C305	&	\phantom{a}		&	\tick	&	\tick	&	\tick	&	\tick	\\
1306$-$09	&	\phantom{a}		&	U	&	--	&	--	&	\tick	&	3C321	&	\phantom{a}		&	\tick	&	low	&	--	&	\tick	\\
1355$-$41	&	\phantom{a}		&	U	&	--	&	--	&	--	&	3C326	&	\phantom{a}		&	--	&	low	&	U	&	--	\\
1547$-$79	&	\phantom{a}		&	U	&	--	&	--	&	--	&	3C382	&	\phantom{a}		&	U	&	--	&	--	&	--	\\
1559$+$02	&	\phantom{a}	3C327	&	--	&	--	&	--	&	--	&	3C388	&	\phantom{a}		&	--	&	--	&	U	&	--	\\
1602$+$01	&	\phantom{a}	3C327.1	&	--	&	--	&	--	&	--	&	3C390.3	&	\phantom{a}		&	U	&	--	&	--	&	--	\\
1648$+$05	&	\phantom{a}	Herc A	&	--	&	U	&	U	&	--	&	3C452	&	\phantom{a}		&	--	&	--	&	--	&	--	\\
1733$-$56	&	\phantom{a}		&	U	&	\tick	&	\tick	&	--	&		&			&		&		&		&		\\
\enddata

\tablecomments{Table showing the RSFA detections for the 2Jy and 3CRR samples. A dash means no evidence for RSFA activity exists for that diagnostic method. Optical column: a tick indicates that the object have young stellar populations identified in careful modeling of optical spectra, U (uncertain) identifies objects where it was not possible to model the spectrum effectively. PAH column: a tick indicates positive strong PAH detection, low indicates a low EW or low S/N PAH detection, while U indicates that good IRS spectra are  not available. Color column: a tick indicates that an object has 70\moo/24\moo$\gtrsim$5, U indicates that the object has an upper limit in 70\moo\ flux. Far-IR column: a tick indicates that the object has a far-IR excess in the sense that it falls 0.5 dex (a factor of 3) above the regression line fitted to $L_{70\mu m}$ versus $L_{[\rm{OIII}]}$ correlation, not including object with upper limits. Note that two of the objects in the 3CRR sample are in common with the 2Jy sample: 3C403 (PKS1949$+$02), 3C445 (PKS2221$-$02). }
\end{deluxetable}

In a study of 33 3C quasars and radio galaxies at intermediate redshifts (0.5$<$z$<$1.1) \citealp{cleary07}  concluded that star formation does not contribute significantly to the MFIR emission. This conclusion was based on  the 15\moo/30\moo\ mid-IR flux ratio also used in other investigations (e.g. \citealp{brandl06}). 

{The study of \citet{shi07} used Spitzer IRS data to investigate PAH emission in 3CR radio galaxies,
as well as other samples of AGN.  Eight objects overlap with the 2Jy and 3CRR samples presented here. The results of \citet{shi07} agree with our study, detecting PAH features in only 2 of the overlapping  objects (3C293, 3C321). The exception is 3C33, where we have identified low EW PAH emission that was not identified in \citet{shi07}.  Moreover, only one object out of the 94 3CR objects included in the \citet{shi07} study has both 7.7 and 11.3\moo\ PAH features detected. Although their sample covers a wider redshift range (z$<$1.5), and their PAH detection method is likely less sensitive than our study, the results shows little evidence for powerful dominant RSFA components in 3CR objects. 

Finally, \citet{fu09} studied a sample of 12 FRII radio-loud quasars with z$\approx$0.3, a relatively similar sample of objects in terms of redshift to the 2Jy sample presented in this paper. They found no evidence for PAH emission in any individual object, or from combined spectra.

Clearly, based on several studies of the PAH features, as well as the other diagnostics discussed in this paper, there is little evidence for powerful RSFA activity in the majority of radio-loud AGN.

We emphasize that, due to the radio flux limited selection method for the two samples, the higher redshift objects are likely to be more powerful radio-loud AGN. It is conceivable that this bias could affect the detection rate of PAH or other RSFA diagnostics across the redshift range of the sample. However, we find no obvious trend with redshift in terms of the rate of detection of RSFA. For example, the rate of detection of RSFA activity in the low redshift 3CRR sample ($z \leq 0.1$: 36\% have optical and/or PAH detection evidence, including low EW PAH detections) is not significantly different from that in the intermediate redshift 2Jy sample ($0.05 < z < 0.7$: 28\%). This confirms that the diagnostic methods we have employed are consistent across the redshift ranges of the two samples.

\subsection{Ratios of PAH emission band fluxes}
\label{sec:ratio}
It has been shown, both experimentally and theoretically, that the ratios between the fluxes of the  6.2,7.7, 8.6 and 11.3\moo\ PAH emission features are sensitive to the radiation environments of the photodissociation regions in which they are produced, as well as the sizes of the aromatic molecules and the degree to which they have been reprocessed in shocks (\citealp{diamond10} and references therein). Since the PAH features are used as a star formation indicator for AGN host galaxies, it is important to investigate whether the PAH emission band spectra are affected by the presence of AGN nuclei. 

In an extensive study of the PAH ratios measured for a large sample of Seyfert galaxies from the Revised Shapley-Ames catalogue (RSA), \citet{diamond10} have found that, on average, the Seyfert galaxies show 7.7/11.3 PAH flux ratios that are low compared with local star forming galaxies, with some Seyfert galaxies identified as having unusually small ratios (7.7/11.3 $<$ 2) -- lower than measured for any of the local star forming HII galaxies in the SINGs sample of \citet{smith07}. This agrees with previous work that also suggested a link between low 7.7/11.3 ratios and the presence of AGN activity (\citealp{smith07}; \citealp{odowd09}).  It is therefore interesting to investigate whether the radio galaxies
in our sample with high EW PAH emission --- which encompass larger AGN luminosities than the \citet{diamond10} sample --- also show low 7.7/11.3 ratios. 

Referring back to Figure \ref{fig_SLpah}, the apparent similarity between PAH spectra of the objects with high EW PAH detections is striking. Considering only the seven objects for which the 6.2, 7.7 and 11.3\moo\ features can be accurately measured (6 objects from the 2Jy sample, and one object from the 3CRR sample) the mean ratios are: 6.2/11.3 =  0.7$\pm$0.2, 7.7/11.3 =  1.7$\pm$0.4,  6.2/7.7 =  0.4$\pm$0.1, see Table \ref{tbl-ratio}.  It is notable that the 7.7/11.3 ratios measured for the radio galaxies fall at the lower end of the range measured for RSA Seyfert galaxies and the Spitzer Infrared Nearby Galaxies Survey (SINGs) HII galaxies; two radio galaxies (PKS0023-16, 3C293) have 7.7/11.3 ratios that are comparable with the most extreme Seyfert galaxies measured by \citet{diamond10}. 

Based on the strong correlation they found between 7.7/11.3 ratios and strong H$_2$S(3) emission, \citet{diamond10} suggested that the low 7.7/11.3 may be due to the reprocessing of the PAH-emitting molecules in shocks, since the rotational H$_2$S(3) feature is enhanced in relatively hot, shocked regions of the ISM (\citealp{ogle10}; \citealp{guillard09}). Although our sample is too small for a thorough statistical investigation, it is notable in that the
objects with the smallest 7.7/11.3 ratio -- PKS0023-26 and 3C293 -- also have the highest ratios of $H_2(S3)$/(11.3+7.7) amongst all the radio galaxies in the combined 2Jy and 3CRR sample (see Table \ref{tbl-ratio}).

In summary, our results on the PAH ratios for the radio galaxies with RSFA are consistent with those obtained for the most extreme RSA Seyferts studied by \citet{diamond10},  reinforcing the view that the 6.6, 7.7 and 8.6\moo\ features are suppressed relative to the 11.3\moo\ feature in the environments of AGN, perhaps as a consequence of reprocessing of the aromatic molecules in shocks. Moreover, following the arguments in \citet{diamond10}, we would also expect the 11.3\moo\ feature to be a good indicator of RSFA in radio galaxies, as it is in other classes of AGN.

\begin{deluxetable}{l@{\hspace{4mm}}c@{\hspace{4mm}}c@{\hspace{4mm}}c@{\hspace{4mm}}l}
\tabletypesize{\scriptsize}
\tablecaption{PAH Ratios\label{tbl-ratio}}
\tablewidth{0pt}
\tablehead{
\colhead{Name}{\hspace{3mm}} & \colhead{6.2/11.3}{\hspace{3mm}} &\colhead{7.7/11.3}{\hspace{3mm}} &\colhead{6.2/7.7} &\colhead{$H_{2}(S3)/(11.3+7.7)$}{\hspace{0mm}} 
}
\startdata									
0023$-$26	&	0.8	$\pm$	0.1	&	1.2	$\pm$	0.1	&	0.7	$\pm$	0.1	&	\phantom{aaaaa}	0.113	$\pm$	0.01	\\
0347$+$05	&	0.7	$\pm$	0.1	&	1.8	$\pm$	0.1	&	0.4	$\pm$	0.1	&	\phantom{aaaaa}	0.008	$\pm$	0.01	\\
0915$-$11	&	0.8	$\pm$	0.1	&	1.8	$\pm$	0.2	&	0.4	$\pm$	0.1	&	\phantom{aaaaa}	0.005	$\pm$	0.01	\\
1733$-$56	&	1.0	$\pm$	0.1	&	1.9	$\pm$	0.2	&	0.5	$\pm$	0.1	&	\phantom{aaaaa}	0.008	$\pm$	0.03	\\
2135$-$20	&	0.5	$\pm$	0.1	&	1.9	$\pm$	0.2	&	0.3	$\pm$	0.1	&	\phantom{aaaaa}	0.018	$\pm$	0.02	\\
2314$+$03	&	0.8	$\pm$	0.1	&	2.4	$\pm$	0.2	&	0.3	$\pm$	0.1	&	\phantom{aaaaa}	0.011	$\pm$	0.01	\\
3C293	&	0.4	$\pm$	0.1	&	1.2	$\pm$	0.1	&	0.3	$\pm$	0.1	&	\phantom{aaaaa}	0.156	$\pm$	0.01	\\
\enddata

\tablecomments{Table presenting the ratios of PAH and $H_2$ fluxes for objects with high EW detections of PAH. See Section \ref{sec:ratio} of the discussion. This table does not include 3C285 and 3C305 that have upper limits for their 6.2 and 7.7\moo\ PAH fluxes. }
\end{deluxetable}

\subsection{The origin of the far-IR emission}
\label{sec:origin}

\begin{figure}[h]
\epsscale{1}
\plotone{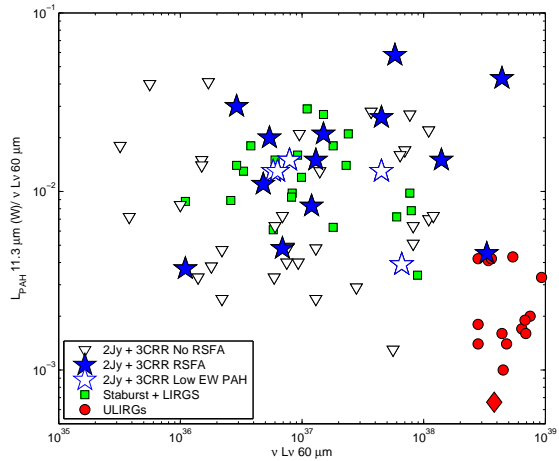}
\caption{: Plot of $L_{11.3\mu m}$/$\nu L_{\nu}$60\moo\ vs $\nu L_{\nu}$60\moo. The 60\moo\ fluxes for the starburst galaxies were obtained from the IRAS catalogue, and the 2Jy and 3CRR sample 70\moo\ fluxes were interpolated to 60\moo\ using the 70/24\moo\ spectral index from the Spitzer data. The IRS data for the starburst galaxies and ULIRGs were downloaded from the Spitzer Heritage Archive and re-reduced and analyzed in an identical way as described for the radio galaxies in this paper. In addition, to compensate for flux loss in the IRS spectra due to the extended structure of the starburst galaxies outside of the slit, a correction factor was applied as discussed in \citet{brandl06}. The ULIRGs sample comprises 15 objects from the sample presented in Veilleux with an average AGN contribution of $<$20\%. The archetypal ULIRG ARP220 is marked with a red diamond. \label{fig_pahFIR} }
\end{figure}

 Based on the correlations between MFIR  and [OIII] luminosities, we previously concluded that the cool, far-IR emitting dust is predominantly heated by AGN illumination for the majority of radio galaxies that show no independent evidence for star formation activity (T07, D09, D10). Under the assumption that that 11.3\moo\ PAH luminosity is a measure of the overall level of star formation activity, if the far-IR continuum emission is dominated by AGN-heated dust, then we expect the ratio of the PAH luminosities to the far-IR continuum luminosities to be lower than measured for pure starburst objects. In this vein, \citet{schweitzer06} found that the $L_{7.7\mu m}$/$\nu L\nu_{60\mu m}$ ratio for PG quasars with detected PAH features ($L_{7.7\mu m}$/$\nu L\nu_{60\mu m}$=0.0110$\pm$0.002) was similar to that of a sample of 12 starburst-dominated ULIRGS ($L_{7.7\mu m}$/$\nu L\nu_{60\mu m}$=0.0130$\pm$0.002), a result that was also discussed by \citet{veilleux09a}. The authors conclude that the similar ratios argue in favor of a starburst origin for the far-IR emission of PG quasars. In contrast, based on similar techniques, \citet{shi07} concluded that the contribution of starburst heating to the far-IR continuum is relatively minor ($\sim$24\%) for most radio-loud AGN.

The problem with using the L$_{\rm{PAH}}$/L$_{\rm{far-IR}}$ ratio as a direct indicator of the degree of starburst heating of the far-IR emitting dust is that the scatter in this ratio is large, even for starburst dominated objects. Results from SINGs have indicated that the use of PAH emission for estimating star formation rates may be uncertain by a factor of $\sim$10 (\citealp{dale05,calzetti07}). This conclusion is also supported by \citet{tacconi05} who find strong variations in the PAH to stellar continuum ration in nearby starburst galaxies, as well as \citet{haas09} who find a range in the $L_{fir}/L_{6.3\mu m}$ of a factor of 10 for a sample of PAH selected galaxies from the ISOCAM parallel survey. In Figure \ref{fig_pahFIR} we plot $L_{11.3\mu m}$/$\nu L_{\nu}$60\moo\ for the 2Jy and 3CRR objects from the  current study, along with  comparison samples of LIRGs and starburst objects of similar far-IR luminosity taken from \citet{brandl06}, and starburst-dominated ULIRGs taken from the study of \citet{veilleux09a}. 

Consistent with the above discussion it is notable that the scatter in the $L_{11.3\mu m}$/$\nu L_{\nu}$60\moo\ ratio is large, both for the radio galaxies we have identified as RSFA objects in our sample, and for the starburst galaxies in the sample of \citet{brandl06}. In addition, and perhaps counter-intuitively, the starburst-dominated ULIRGs tend to the lowest values of $L_{11.3\mu m}$/$\nu L_{\nu}$60\moo. An explanation for this difference might be that a significant fraction of the starburst activity in the ULIRGs is so deeply embedded and extinguished by dust that it is not detected using the short wavelength PAH features (see Veilleux et al. 2009 for further discussion).
 
Given the high degree of scatter and large number of relatively high upper limits in the plot, it is difficult to make strong conclusions about the heating mechanism for the far-IR emitting dust for most of the radio galaxies in our sample based on this evidence alone. However, the upper limits for some radio galaxies fall well below the majority of values measured for the Brandl et al. LIRG and starburst galaxies ($L_{11.3\mu m}$/$\nu L_{\nu}$60\moo\ $<$ $5\times10^{-3}$), suggesting that AGN illumination makes a large contribution to the heating of the cool dust in these objects, supporting our previous conclusions (T07, D09). It is noteworthy that many of the objects with $L_{PAH11.3\mu m}$/$\nu L_{\nu}$ 60\moo\ $<$ $5\times10^{-3}$ comprise the low-z NLRG with high S/N spectra (e.g. PKS1559+02, PKS1949+02, PKS2250-41, PKS2356-61) as well as low-z BLRG (3C382, 3C390.3) with the very highest S/N spectra. This suggests that many of the objects with low S/N spectra (mainly high-z objects) might fall in this region of the diagram if they had better S/N data.

These results are relevant to the interpretation of the far-IR emission from AGN in the distant Universe.
High redshift galaxies that show evidence for AGN activity at X-ray \citep{alexander05}, optical \citep{priddey03} and radio (\citealp{archibald01}; \citealp{willott02}) wavelengths have been detected with deep surveys at sub-mm wavelengths which are sensitive to the redshifted far-IR emission from cool dust.  Because investigations of starbursts galaxies in the local universe have revealed prodigious far-IR radiation, the sub-mm emission has been interpreted as being due to cold dust heated by strong star formation activity. Our previous work on the 2Jy sample (T07, D09, D10) has shown that this interpretation is not unique because the dust emitting the thermal mid- to far-infrared (MFIR) radiation can be heated by either starbursts or by  AGN. Figure \ref{fig_pahFIR} provides evidence that, in many powerful radio galaxies, far-infrared emission is dominated by AGN and not starburst heating. Therefore, our results suggest that the interpretation of sub-mm emission from distant radio galaxies as originating purely from star formation should be treated with caution.

\subsection{Compact radio sources}
\label{sec:css}

Compact steep spectrum (CSS) and gigahertz peaked sources (GPS) are characterized by the small sizes of their radio  structures (D $<$ 30 kpc and $<$1kpc respectively) and steep high frequency radio spectra. Mounting evidence supports the idea that CSS/GPS objects represent a young phase of more extended powerful radio-loud AGN (\citealp{fanti95}, \citealp{readhead96}). In particular, the measured proper motions of the radio components in some CSS objects that imply very young ages ($<$3$\times10^{3}$ years; e.g. \citealp{polatidis03}). 

The combined 2Jy and 3CRR sample includes 7 CSS sources and 1 GPS  source (PKS1934-63). Three of these objects have high equivalent with PAH detections (PKS0023-26, 3C305, PKS2135-20) and three additional objects have been identified with low EW PAH detections (PKS1151-34, PKS1814-63, PKS1934-63). The rate of CSS/GPS objects in the 2Jy+3CRR sample with PAH detections is therefore 75\% (6/8) --- much higher than the detection rate for the extended radio sources in the 2Jy+3CRR sample with good IRS spectra (21\%: 10/48). Interestingly, similar results have been obtained by \citet{willett10}, who detect PAH emission in  7/8 (88\%) of the objects in their heterogeneous sample of nearby compact symmetric objects\footnote{The CSOs represent a subset of the compact radio source population, with relatively symmetric radio structures; most of the compact (CSS,GPS) sources in the 2Jy+3CRR sample would also be classified as CSOs.} ($z < 0.26$). This evidence for an enhanced degree of star formation in the host galaxies of compact radio sources is also consistent with the fact that the compact sources are disproportionately represented in the (small) subset of radio galaxies that show strong evidence for young stellar populations in their optical spectra \citep{tadhunter11}. 

At first sight it might be thought that the relatively high incidence of starburst activity among the compact radio sources is connected with the relative youthfulness of such sources: if the radio source activity is triggered close to the peak of the starburst associated with the triggering events, then the compact sources might be expected to show enhanced star formation activity compared to the more extended radio sources
that are observed at later epochs. However, the problem with this explanation is that both models (\citealp{mihos96,dimatteo07}) and direct observations \citep{rodriguez10} suggest that the main starburst episodes associated with major galaxy mergers are relatively long-lived: $\sim$10---100~Myr -- much longer than the expected time period over which a radio source would appear as a compact radio source, but similar to expected lifetimes of extended radio sources. Moreover, the spectroscopic signatures of the merger-induced starburst would be detectable in optical spectra for several hundred Myr after the starburst. Therefore we would not expect to find a major difference between the rate of detection of starburst signatures in compact and extended radio sources, given the similarities between the expected timescales of (extended) radio sources and starburst activity.

Alternatively,  \citet{tadhunter11} have suggested that the association between starburst activity and compact radio sources may be a consequence of an observational selection effect: young radio sources triggered in major gas-rich mergers, that also lead to starbursts, will have their radio emission enhanced by jet/cloud interactions as their radio jets plough through the dense merger debris. This will lead to such sources being preferentially selected in flux limited radio samples such as the 2Jy and 3CRR, boosting their numbers relative to the radio sources that are not triggered in major gas-rich mergers; the compact radio sources triggered in this way will have a higher ratio of radio power to intrinsic jet and AGN power than more extended radio sources triggered in less dense environments. As these compact sources expand further into the more tenuous haloes of the host galaxies their radio fluxes will decline and they will drop out of the flux limited samples. We note that this selection effect is consistent with the fact that the compact radio sources tend to fall below the correlations between [OIII] and 24$\mu$m emission and radio power, as expected if the radio flux is boosted relative to the intrinsic AGN power (\citealp{morganti97}; \citealp{holt09}; Morganti et al. 2011). It may also help explain why \citet{odea97} found that compact radio sources are more common in flux limited radio samples than would be
expected on the basis of a simple extrapolation of the radio size vs. number relation
for more extended radio sources}\footnote{Whereas the extended radio sources ($D > 10$~kpc) show a rapid increase in number within bins of increasing linear size for the radio sources, the compact radio sources ($D < 10$~kpc) show a constant number with size --- there are far more compact sources than expected on the basis of a simple extrapolation of the number vs size relationship for the extended sources \citep[see][Figure 10]{odea97} towards small sizes. }

\subsection{Consequences for the triggering and evolution of radio source host galaxies}

In parallel work, we have found that at least 85\% of the objects in the 2Jy sample show morphological evidence for recent galaxy interactions  and mergers \citep{ramos11a,ramos11b}. However, the morphologies and stellar population properties displayed by the radio source host galaxies are diverse and do not all correspond to a single phase of a particular type of galaxy interaction (\citealp{ramos11a,ramos11b,tadhunter11}). While some radio sources have indeed clearly been triggered at the peaks of major gas-rich mergers, others are seen at a relatively late post-merger phases well {\it after} the coalescence of the merging nuclei; and a significant subset show tidal links with well-separated companion galaxies, suggesting that they have been triggered after the first pass of the interacting galaxies, but well {\it before} the final merger (if any). These results indicate that it is
possible for the gas inflows associated with a range of stages and types of galaxy interactions to be sufficient to trigger powerful radio galaxies, many of which are associated with AGN of quasar-like
luminosity ($L_{BOL} > 10^{38}~W$). Overall, these results are inconsistent with many recent hydrodynamical simulations which predict that the most powerful, quasar-like AGN activity is only triggered around the time of coalescence of the merging nuclei in major, gas-rich mergers (e.g. \citealp{dimatteo05}), or with a short time delay \citep{hopkins11}. Our mid-IR results on the PAH features strongly reinforce these conclusions: in most objects we do not detect the levels of RSFA that would be expected if all powerful radio galaxies were triggered at the peaks of major, gas-rich mergers\footnote{Note that we find no clear link between those objects we have identified in this study with evidence for RSFA and any one particular stage or type of interaction (i.e. with signatures of pre-coalescence, coalescence and post-coalescence) that have been identified in the host galaxy morphologies.} We suggest that the current hydrodynamical simulations do not adequately capture the physics of the gas flows across the variety of interaction types and/or stages that lead to the triggering of radio-loud AGN. 

It is also important to consider whether the negative feedback effect of the AGN and jet activity might be responsible for the low levels of RSFA detected in radio galaxies i.e. whether the AGN and jets in these powerful radio galaxies input sufficient energy into the ISM of the host galaxies to suppress some of the star formation associated with the triggering events. Because star formation associated with advanced mergers is often concentrated near the nucleus, the energy input of the AGN may have a major impact on these star formation regions. Given that we have found evidence for RSFA in up to 35\% of our samples, it appears that the AGN cannot entirely shut down star formation in powerful radio galaxies. On the other, we note that two of the objects in this investigation (3C293, 3C326) have been associated with AGN feedback processes through detailed observations of the molecular gas content. The study of \citet{nesvadba10} found evidence that these objects have star formation rates lower than expected for the amount of dense gas available for star formation.

The feedback effect of the AGN may be significant in some objects,
however,
the lack of evidence for RSFA in the 2Jy and 3CRR samples may simply be
due to the fact that we are observing the sources a long time before or
after any main merger-induced starburst. Alternatively, the triggering
galaxy interactions may not be of the types that are associated with major
star formation episodes (e.g. they may be relatively minor or gas-poor).
Clearly, in order to develop our understanding of the triggering
events further, we require  information about
the total reservoir of cool ISM in the host galaxies, as
well as the details of  the circum-nuclear gas kinematics. 

\section{Conclusions}

We have presented Spitzer/IRS spectroscopy for complete samples of 2Jy and 3CRR radio-loud AGN, detecting 93\% and 79\% of the objects in the two samples respectively. Our analysis of the spectra reveals strong RSFA-tracing PAH features in only a minority of the objects from the two samples (16\%) that have good IRS spectra. 
Combining this result with optical continuum spectroscopy, mid- to far-IR (MFIR) color and far-IR excess diagnostics, we find that only 35\%  of objects in the combined 2Jy and 3CRR sample show any evidence for RSFA at optical and/or MFIR wavelengths.
 \emph{This result argues strongly against the idea that there is a close link between starbursts and powerful radio-loud AGN, reinforcing the view that only a minority are triggered at the peaks of star formation activity in major, gas-rich mergers.}
The PAH emission also allows us to test whether there exists a substantial proportion of RSFA that is obscured by dust that was undetected at optical wavelengths. Although we do find significant PAH emission in  4 objects that were not previously identified as hosting RSFA on the basis of other star formation indicators, this does not substantially change the statistics for the overall rate of detection of RSFA in radio galaxies from our previous investigations. 
In addition, we find that compact radio sources show a significantly higher incidence of RSFA, which cannot be readily explained by the youthful nature of these objects. We suggest that, for radio selected samples, there may be a bias towards the selection of compact radio sources that are triggered in gas-rich environments.

\acknowledgments We would like to thank Jack Gallimore for assistance with the modification of PAHFIT. We would also like to thank the anonymous referee for their comments which have improved this investigation. This work is based [in part] on observations made
with the Spitzer Space Telescope, which is operated by the Jet
Propulsion Laboratory, California Institute of Technology under a
contract with NASA. This research has made use of the NASA/IPAC
Extragalactic Database (NED) which is operated by the Jet Propulsion
Laboratory, California Institute of Technology, under contract with
the National Aeronautics and Space Administration. 
D. D. acknowledges support from NASA grant based on observations from Spitzer program 50588 and the NASA ROSES ADAP program. M.B.N.K. was supported by the Peter and Patricia Gruber Foundation through the IAU-PPGF fellowship, by the Peking University One Hundred Talent Fund (985), and by the National Natural Science Foundation of China (grants 11010237 and 11043007). C.R.A. acknowledges financial support from STFC PDRA (ST/G001758/1). KJI is funded through the Emmy Noether Programme of the German Science Foundation (DFG).

{\it Facilities:} \facility{Spitzer (IRS)}

\bibliographystyle{apj.bst} 
\bibliography{bib_list}

\appendix

\begin{deluxetable}{l@{\hspace{0mm}}l@{\hspace{0mm}}c@{\hspace{0mm}}c@{\hspace{0mm}}}
\tabletypesize{\scriptsize}
\tablecaption{Ancillary data \label{tbl-appendix}}
\tablewidth{0pt}
\tablehead{
\colhead{Name}{\hspace{0mm}} &\colhead{z}{\hspace{-3mm}} &\colhead{$S_{11.3}$(W/$m^2$)}{\hspace{0mm}} & \colhead{$S_{60}$(W/Hz)}{\hspace{-3mm}} 
}
\startdata
\cutinhead{ULIRGs}								
F12112+0305	\phantom{aaa}	&	0.0733	&	8.5E-16	&	2.5E+26	\\
F17207-0014	\phantom{aaa}	&	0.0428	&	3.5E-15	&	2.9E+26	\\
F20414-1651	\phantom{aaa}	&	0.0871	&	3.6E-16	&	1.8E+26	\\
F13335-2612	\phantom{aaa}	&	0.1250	&	3.5E-16	&	1.3E+26	\\
F03250+1606	\phantom{aaa}	&	0.1290	&	3.5E-16	&	1.4E+26	\\
F10565+2448	\phantom{aaa}	&	0.0431	&	2.8E-15	&	1.1E+26	\\
F23234+0946	\phantom{aaa}	&	0.1279	&	1.7E-16	&	1.7E+26	\\
F14348-1447	\phantom{aaa}	&	0.0830	&	7.6E-16	&	2.6E+26	\\
F22491-1808	\phantom{aaa}	&	0.0778	&	3.1E-16	&	1.7E+26	\\
F16333+4630	\phantom{aaa}	&	0.1910	&	3.0E-16	&	3.4E+26	\\
Arp 220	\phantom{aaa}	&	0.0181	&	3.5E-15	&	1.6E+26	\\
F14197+0813	\phantom{aaa}	&	0.1310	&	1.1E-16	&	1.1E+26	\\
F05024-1941	\phantom{aaa}	&	0.1920	&	1.1E-16	&	2.6E+26	\\
F01494-1845	\phantom{aaa}	&	0.1580	&	3.5E-16	&	2.0E+26	\\
F00482-2721	\phantom{aaa}	&	0.1292	&	9.3E-17	&	1.1E+26	\\
\cutinhead{LIRGS and starbursts}							
IC 342...	\phantom{aaa}	&	0.0001	&	1.8E-14	&	8.3E+21	\\
Mrk 52...	\phantom{aaa}	&	0.0071	&	2.1E-15	&	1.1E+24	\\
Mrk 266...	\phantom{aaa}	&	0.0155	&	2.7E-15	&	7.9E+24	\\
NGC 520...	\phantom{aaa}	&	0.0076	&	2.9E-14	&	8.1E+24	\\
NGC 660...	\phantom{aaa}	&	0.0028	&	4.9E-14	&	2.3E+24	\\
NGC 1097...	\phantom{aaa}	&	0.0042	&	4.3E-14	&	4.2E+24	\\
NGC 1222...	\phantom{aaa}	&	0.0082	&	6.5E-15	&	3.8E+24	\\
NGC 1365...	\phantom{aaa}	&	0.0055	&	3.0E-14	&	1.2E+25	\\
NGC 1614...	\phantom{aaa}	&	0.0159	&	1.6E-14	&	3.6E+25	\\
NGC 2146...	\phantom{aaa}	&	0.0030	&	1.6E-13	&	5.7E+24	\\
NGC 2623...	\phantom{aaa}	&	0.0185	&	4.2E-15	&	3.7E+25	\\
NGC 3256...	\phantom{aaa}	&	0.0094	&	4.1E-14	&	4.0E+25	\\
NGC 3310...	\phantom{aaa}	&	0.0033	&	1.6E-14	&	1.7E+24	\\
NGC 3556...	\phantom{aaa}	&	0.0023	&	3.0E-14	&	7.7E+23	\\
NGC 3628...	\phantom{aaa}	&	0.0028	&	3.6E-14	&	1.9E+24	\\
NGC 4088...	\phantom{aaa}	&	0.0025	&	1.9E-14	&	7.5E+23	\\
NGC 4194...	\phantom{aaa}	&	0.0083	&	1.6E-14	&	7.1E+24	\\
NGC 4676...	\phantom{aaa}	&	0.0220	&	3.7E-15	&	5.9E+24	\\
NGC 4818...	\phantom{aaa}	&	0.0036	&	8.9E-15	&	1.1E+24	\\
NGC 4945...	\phantom{aaa}	&	0.0019	&	1.9E-13	&	9.7E+24	\\
NGC 7252...	\phantom{aaa}	&	0.0160	&	6.0E-15	&	4.6E+24	\\
NGC 7714...	\phantom{aaa}	&	0.0093	&	7.0E-15	&	4.3E+24	\\
\enddata

\tablecomments{Table presenting the ancillary data from Figure \ref{fig_pahFIR} for ULIRG, LIRG and starburst objects. }
\end{deluxetable}

\end{document}